\documentclass[twocolumn]{aastex63}

\usepackage{xcolor}

\definecolor{my_color}{HTML}{3a18b1}

\newcommand{\cannon}{\texttt{TheCannon}}
\newcommand{\teff}{\mbox{$T_{\rm eff}$}}

\newcommand{\feh}{\mbox{$\rm [Fe/H]$}}
\newcommand{\mg}{\mbox{$\rm [Mg/Fe]$}}

\newcommand{\alphafe}{\mbox{$\rm [\alpha/Fe]$}}

\newcommand{\logg}{\mbox{$\log g$}}

\newcommand{\TSun}{\ifmmode {T_{\odot}}\else${T_{\odot}}$\fi}
\newcommand{\LSun}{\ifmmode {\logg_{\odot}}\else${\logg_{\odot}}$\fi}
\newcommand{\FSun}{\ifmmode {\feh_{\odot}}\else${\feh_{\odot}}$\fi}
\newcommand{\mSun}{\ifmmode {M_{\odot}}\else${M_{\odot}}$\fi}

\usepackage{appendix}
\usepackage{amsmath}
\usepackage{mathtools}

\shorttitle{Gaia RVS Solar Analog Abundances}
\shortauthors{Rampalli et al.}

\begin{document}

\title{The Sun Remains Relatively Refractory Depleted: Elemental Abundances for 17,412 Gaia RVS Solar Analogs and 50 Planet Hosts}

\correspondingauthor{Rayna Rampalli}
\email{raynarampalli@gmail.com}

\author[0000-0001-7337-5936]{Rayna Rampalli}
\altaffiliation{NSF GRFP Fellow} 
\affiliation{Department of Physics and Astronomy, Dartmouth College, Hanover, NH 03755, USA}

\author[0000-0001-5082-6693]{Melissa K. Ness}
\affiliation{Research School of Astronomy \& Astrophysics, Australian National University, Canberra, ACT 2611, Australia}
\affiliation{Department of Astronomy, Columbia University, 550 West 120th Street, New York, NY, 10027, USA}
\affiliation{Center for Computational Astrophysics, Flatiron Institute, 162 Fifth Avenue, New York, NY 10010, USA}

\author[0000-0002-3285-4858]{Graham H. Edwards}
\affiliation{Department of Earth Sciences, Dartmouth College, Hanover, NH 03755, USA}
\affiliation{Department of Physics and Astronomy, Dartmouth College, Hanover, NH 03755, USA}

\author[0000-0003-4150-841X]{Elisabeth R. Newton}
\affiliation{Department of Physics and Astronomy, Dartmouth College, Hanover, NH 03755, USA}

\author[0000-0001-9907-7742]{Megan Bedell}
\affiliation{Center for Computational Astrophysics, Flatiron Institute, 162 Fifth Avenue, New York, NY 10010, USA}




\begin{abstract}

The element abundances of stars, particularly the refractory elements (e.g., Fe, Si, Mg), play an important role in connecting stars to their planets. Most Sun-like stars do not have refractory abundance measurements since obtaining a large sample of high-resolution spectra is difficult with oversubscribed observing resources. In this work we infer abundances for C, N, O, Na, Mn, Cr, Si, Fe, Ni, Mg, V, Ca, Ti, Al, and Y for solar analogs with Gaia RVS spectra (R=11,200) using the Cannon, a data-driven method. We train a linear model on a reference set of 34 stars observed by Gaia RVS with precise abundances measured from previous high resolution spectroscopic efforts (R $> 30,000-110,000$). We then apply this model to several thousand Gaia RVS solar analogs. This yields abundances with average upper limit precisions of 0.04--0.1 dex for 17,412 stars, 50 of which are identified planet (candidate) hosts. We subsequently test the relative refractory depletion of these stars with increasing element condensation temperature compared to the Sun. The Sun remains refractory depleted compared to other Sun-like stars regardless of our current knowledge of the planets they host. This is inconsistent with theories of various types of planets locking up or sequestering refractories. Furthermore, we find no significant abundance differences between identified close-in giant planet hosts, giant planet hosts, and terrestrial/small planet hosts and the rest of the sample within our precision limits. This work demonstrates the utility of data-driven learning for future exoplanet composition and demographics studies.

\end{abstract}

\keywords{stars: abundances, stars: solar-type - sun: abundances – techniques: spectroscopic – exoplanet: formation - solar system: formation}
\submitjournal{ApJ}
\accepted{February 23, 2024}
\section{Introduction}\label{sec:intro}

Know thy star, know thy planet: planets can reflect chemical properties of their host star because they are formed from the same molecular cloud. Studying star-planet connections of other planetary systems using their host star chemistry can answer open questions we have about our own solar system. 
The Sun shows a trend of relative depletion in refractory elements\footnote{We define an element as ``refractory'' if its 50\% condensation temperature from \cite{Lodders03} $> 900$ K following \cite{Flores23}.} (Na, Mn, Cr, Si, Fe, Ni, Mg, V, Ca, Ti, Al, Y) with increasing condensation temperature compared to 80\% of its Sun-like counterparts \citep{bedell18}, but the source of this relative depletion is unknown. Early work from \cite{Melendez09,Ramirez09} posits that the terrestrial planets have locked up these refractory elements. More recently, \cite{Booth20} suggested that giant planets (e.g. Jupiter) can create dust traps preventing the infall of refractory rich dust onto the host star. In this work we aim to determine if this apparent refractory depletion is due to the planets in the solar system by comparing refractory abundances for solar analogs and identified solar analog planet hosts.


When broadly considering the relationship between host star chemistries and their planets, there is a correlation between planet occurrence and host star iron abundance (\feh\ as a proxy for bulk metallicity) through exoplanet occurrence rate and demographics studies. \cite{Fischer05} demonstrated that giant planets are more commonly found around high metallicity stars, but there is a lack of consensus on whether the same relationship holds for smaller terrestrial planets \citep{Buchhave14,Petigura18}. Enhancement in abundances of individual elements generally increases with planet occurrence as well \citep{Wilson22}.


Thus, precise abundances for planet hosts are a necessity to understanding their planets' role in shaping their compositions and vice versa. Individual planet characterization efforts require observations from ground-based telescopes, and obtaining follow-up high resolution spectroscopy for each system is difficult due to instrument oversubscription. Most Sun-like stars do not have precise abundance measurements, including for the refractory elements. This has limited opportunities to understand the connections between host star and planet chemical compositions on a demographic scale. However, the advent of large-scale spectroscopic and astrometric surveys of stars in our Galaxy and resulting data-driven methods complement planet characterization efforts quite well. The Gaia mission \citep{gaiamission} alone has measured precise astrometry for over 1.4 billion stars. In its third data release \citep{GaiaDR3}, just under 1 million medium-resolution spectra were released with 149 million more expected in the coming years from the Radial Velocity Spectrometer (RVS) instrument \citep{recio-blanco}. 

To investigate whether the source of the Sun's refractory depletion is due to its status as a planet host, we infer abundances for several thousand solar analogs (\teff: $\TSun\pm 300$ K, \logg: $\LSun\pm 0.3$, \feh: $\FSun\pm 0.3$ dex, as defined by \citealt{Berke23}) observed by Gaia, a subset of which are identified planet hosts. We compile abundance information for solar analogs from archival high resolution spectroscopic surveys that also have observed Gaia RVS spectra. We call this our \textit{reference set}. We then train a linear model with \cannon\ \citep{cannon} on these spectra and abundances, apply it to other Gaia stars (our \textit{test set}) to infer precise abundances, and examine refractory depletion trends with respect to the Sun and between planet host populations.  

In Section \ref{sec:data}, we discuss the literature solar analogs used to make up our reference set, the Gaia RVS spectra used to make up our test set, and the planet host catalogs we crossmatch with our test set of stars. We outline what \cannon\ is, how it is used to test the robustness of the reference set and the model, how it is applied to the test set, and how we inferred isochrone ages for the test set in Section \ref{sec:methods}. In Section \ref{sec:results}, we show \cannon\ results and the resulting refractory depletion trends with the test set. We also show the precision limits Gaia RVS spectra allow. We conclude with a discussion of our results in Section \ref{sec:conclusions}.

\section{Data} \label{sec:data}

\subsection{Literature Solar Analogs}\label{sec:lsa}
To build our reference set sample of stars, we search the literature for solar analogs (\TSun: $5772 \pm 300$ K, \LSun: $4.44 \pm 0.3$, \FSun: $0.0 \pm 0.3$ dex as defined by \citealt{Prvsa16}) with $
\geq 10$ precise abundances measured from high-resolution spectra. We find 1294 unique stars from the \cite{Bensby14,Hinkel14,Brewer16,Brewer18,bedell18} catalogs with abundance measurements for C, N, O, Na, Mn, Cr, Si, Ni, Mg, V, Ca, Ti, Al, and Y. 

\cite{Bensby14} conducted a high-resolution (R=40,000--110,000) spectroscopic survey of 714 F and G dwarf and subgiant stars in the Solar neighborhood. This was done using the FEROS spectrograph on the ESO 1.5 m and 2.2 m telescopes, the SOFIN and FIES spectrographs on the Nordic Optical Telescope, the UVES spectrograph on the ESO Very Large Telescope, the HARPS spectrograph on the ESO 3.6 m telescope, and the MIKE spectrograph on the Magellan Clay telescope. Average errors are 56 K for \teff, 0.08 dex for \logg, and $< 0.1$ dex for [X/Fe] which we convert to [X/H]. \cite{Hinkel14} compiled spectroscopic abundance data from 84 literature sources for 50 elements across 3058 stars in the solar neighborhood, within 150 pc of the Sun. The different instruments used are more than we can feasibly list here, but the minimum resolution reported starts at R=30,000. Average errors are 100 K for \teff\ 
and 0.03--0.1 dex for [X/H]. \cite{Brewer16} and \cite{Brewer18} performed a uniform spectroscopic analysis of $ \sim 2,700$ F, G, and K dwarfs through the California Kepler Survey with Keck HIRES (R=70,000). Average errors are 27 K for \teff, 0.05 dex for \logg, and $< 0.1$ dex for [X/H]. \cite{bedell18} conducted a spectrospic study and measured 30 elemental abundances for 79 Sun-like stars within 100 parsecs using the HARPS spectrograph (R=115,000) on the 3.6 meter telescope of ESO. These stars had the smallest average errors with 4 K for \teff, 0.01 dex for \logg, and $< 0.02$ dex for [X/H].

We use the respective catalog ID names and crossmatch them to Gaia stars using the Gaia Archive with a two arcsecond radius. Of these 1294 stars, only 105 had observed Gaia RVS spectra available in Gaia DR3 (11 from \citealt{Bensby14}, 11 from \citealt{Hinkel14}, 78 from \citealt{Brewer16,Brewer18}, and 5 stars from \citealt{bedell18}). Upon examining the RVS spectra of the literature stars individually, we find a handful of spectra that deviate from a typical Sun-like spectrum such as magnetically-induced active stars showing emission at the Ca triplet \citep{Martin17}. Because we are interested in finding solar \textit{analogs}, we want to exclude such outliers from our eventual reference set. We construct a ``median solar analog spectrum'' by choosing the median RVS flux value at each wavelength step across the entire literature sample. The median solar analog spectrum is then compared to each individual literature star's spectrum using two $\chi^{2}$ tests. The spectra here extend from 846.4 to 869.6 nm in steps of 0.01 nm yielding 2321 wavelength steps. We compare across I) the entire spectrum and II) at the calcium triplet lines as this is where the most variance is usually found \citep{Rampalli21}. We define a reasonable\footnote{A $\chi^{2} < $ the number of degrees of freedom indicates the data are being overfitted, and a $\chi^{2} > $ 3--5 times the degrees of freedom indicate the data are not being fit well. We choose 2 to be conservative.} $\chi^{2}$ as $\leq$ two times the degrees of freedom, which corresponds to I) the number of wavelength steps making up the entire spectrum (2321) and II) the number of wavelength steps containing the calcium triplet lines (69). We remove the stars that do not meet this $\chi^{2}$ threshold. We then reconstruct the median solar analog with the updated set of stars and repeat the $\chi^{2}$ comparison. In this second iteration, we do not find any stars that have an unreasonably high $\chi^{2}$. We are left with 96 stars, and these are the stars that make up our reference set. None of the spectroscopic outliers we remove have unique abundance patterns that deviate from the abundance distributions of our finalized reference set.

 While four catalogs have measurements for C, N, Mn, and V, \cite{Bensby14} does not. For our analysis with \cannon, we initially take the mean of the other catalogs' measurements for each of these elements and use these mean values for the Bensby stars. We do the same for the \cite{bedell18} catalog, which did not report abundances for N. Ultimately, we choose to exclude stars from \cite{Bensby14} to improve our inferred abundance precision. We describe this decision further in Section \ref{sec:refset}. This decreases our reference set from 96 stars to 85 stars though we do not make this cut \textit{until} we use \cannon\ to infer abundances. We still include the Bensby stars when deciding what our test set of stars will be.

\begin{figure*}
    \centering
    \includegraphics[width=0.85\textwidth]{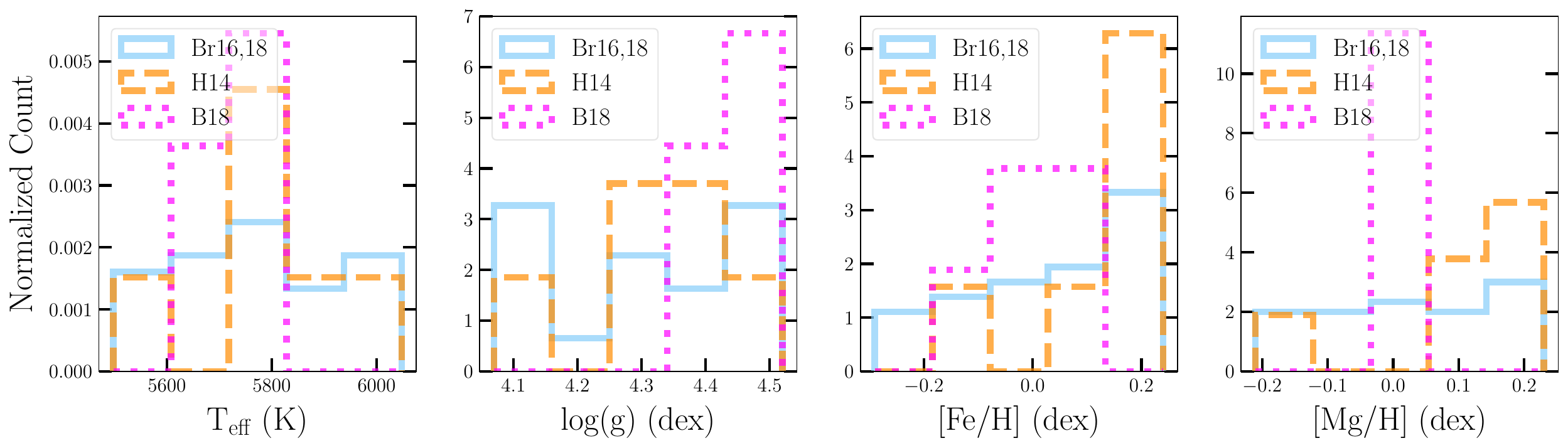}
    \includegraphics[width=0.85\textwidth]{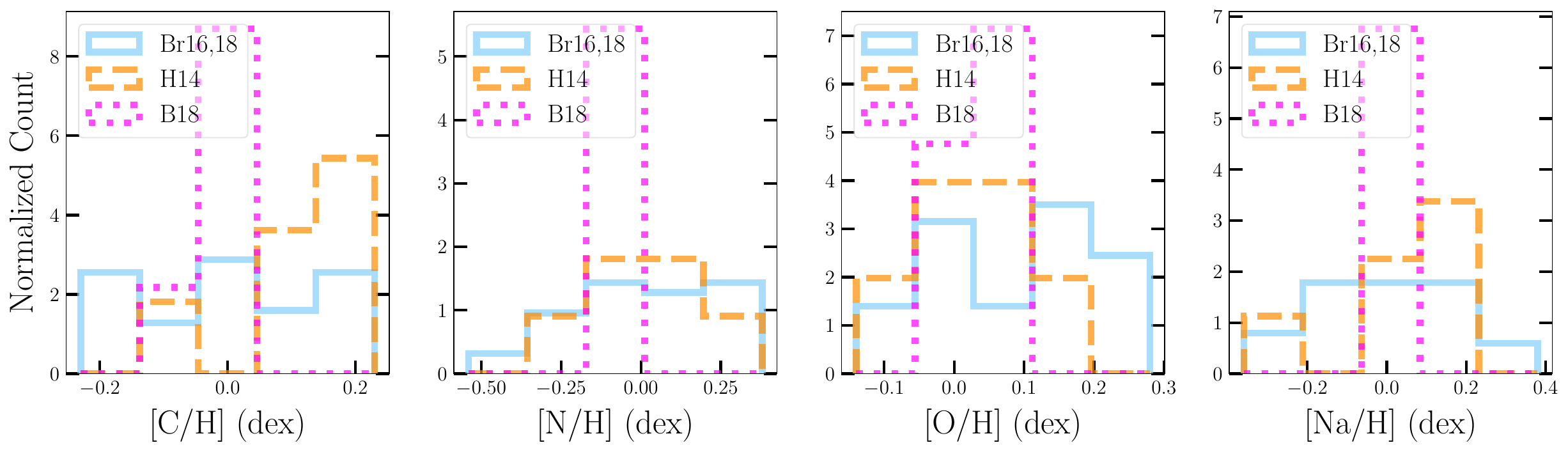}
    \includegraphics[width=0.85\textwidth]{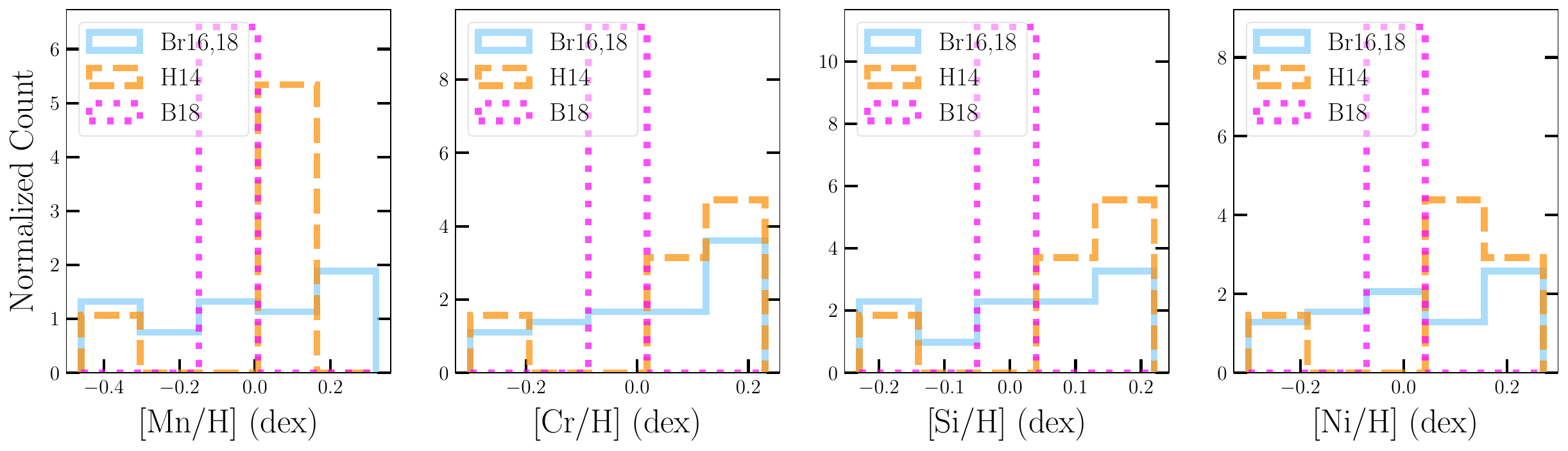}
    \includegraphics[width=0.85\textwidth]{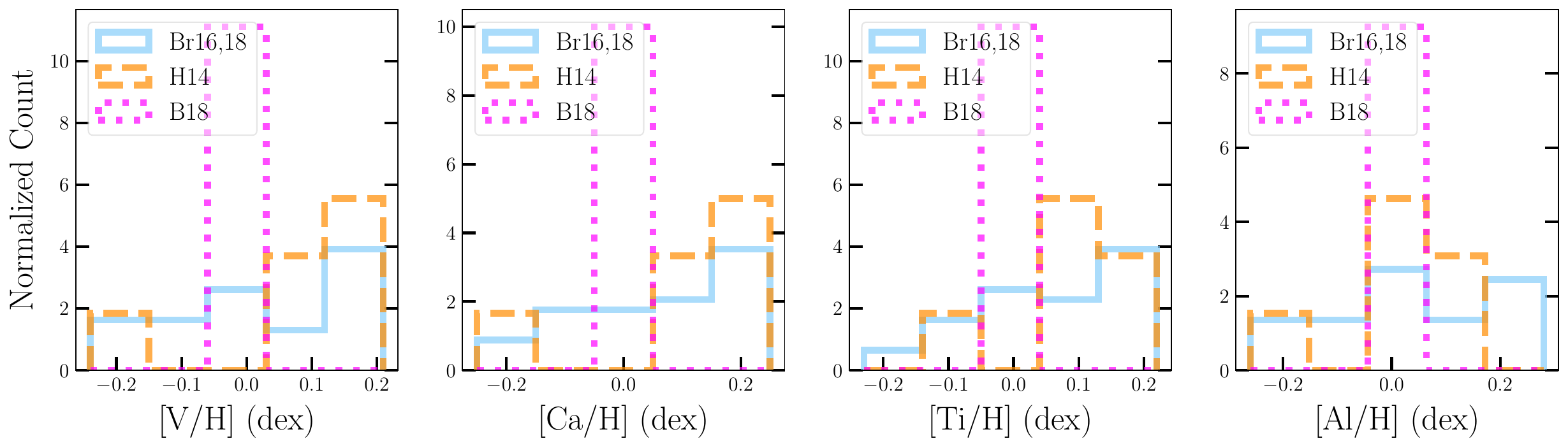}
    \includegraphics[width=0.85\textwidth]{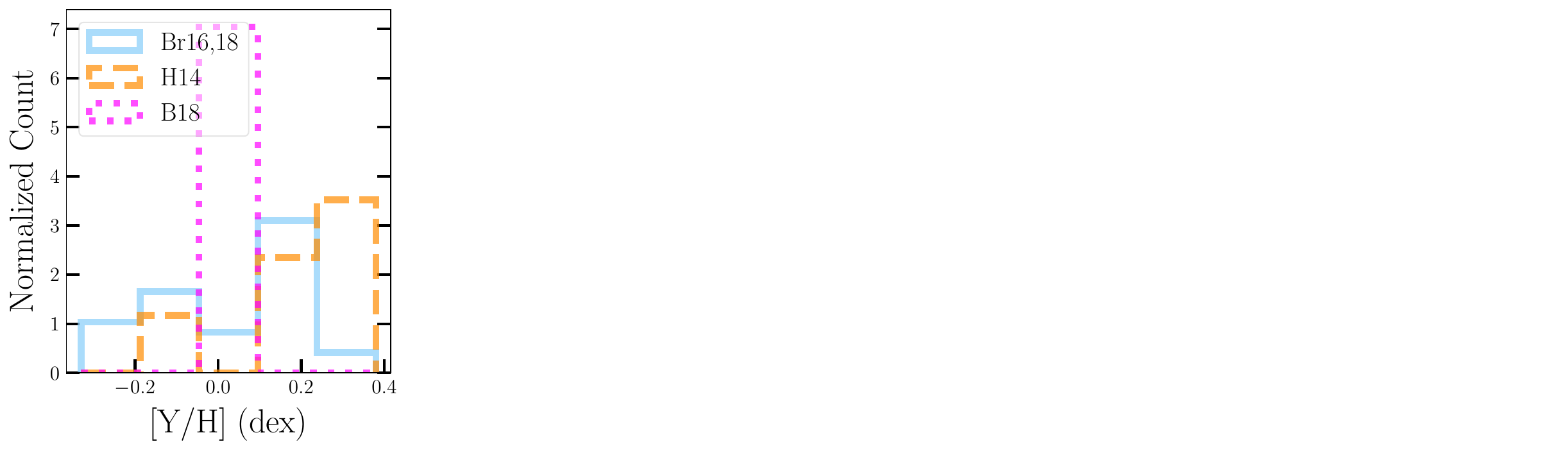}

    \caption{Distributions in \teff, \logg, Fe, Mg, C, N, O, Na, Mn, Cr, Si, Ni, V, Ca, Ti, Al, and Y with respect to H for reference set. \cite{Brewer16,Brewer18} or Br16,18 in light blue, 
    \cite{Hinkel14} or H14 in orange, and \cite{bedell18} or B18 in pink.}
    \label{fig:rset_labels}
\end{figure*}

 We show the reference set's distributions in \teff, \logg, and elements Fe, Mg, C, N, O, Na, Mn, Cr, Si, Ni, V, Ca, Ti, Al, and Y with respect to H in Figure \ref{fig:rset_labels}. \cite{Brewer16,Brewer18} stars are light blue and labeled as Br16,18, \cite{Hinkel14} stars are orange and labeled as H14, and \cite{bedell18} stars labeled as B18 are pink. The distributions of stars from each catalog are wide in comparison to those from \cite{bedell18} since the latter specifically targeted solar twins. We show the stars' Gaia DR3 names, their coordinates, literature measurements, the signal-to-noise (S/N) of their Gaia RVS spectra, and a selection of the parameters measured from the RVS spectra in Table \ref{t:rs}. 

\begin{deluxetable*}{lcccl}
\tablecaption{Reference Set Parameters \label{t:rs}}
\tablewidth{0pt}
\tablehead{
\colhead{Column} & \colhead{Format} & \colhead{Units} & \colhead{Example} & \colhead{Description}
}
\startdata
\sidehead{\textit{Identifier:}} 
DR3id  & integer & \nodata & 746545172372256384 & Gaia DR3 ID \\
\sidehead{\textit{Coordinates:}} 
ra      & float & degrees & 150.249968 & Right ascension \\
dec     & float & degrees & 31.921762 & Declination \\
\sidehead{\textit{Literature Values:}} 
teff   & integer  & K    & 5742 & \teff\ measurement \\
logg   & float  & dex    & 4.31 & \logg\ measurement  \\
Fe\_H\_     & float  & dex    & 0.2 & \feh\ measurement  \\
Mg\_H\_  & float  & dex    & 0.21 & [Mg/H] measurement \\
C\_H\_     & float  & dex     & 0.15 & [C/H] measurement \\
N\_H\_    & float  & dex     & 0.13 & [N/H] measurement \\
O\_H\_   & float  & dex & 0.18 & [O/H] measurement \\
Na\_H\_    & float  & dex & 0.23 & [Na/H] measurement\\ 
Mn\_H\_    & float  & dex & 0.23 & [Mn/H] measurement \\
Cr\_H\_    & float  & dex & 0.2 & [Cr/H] measurement  \\
Si\_H\_    & float  & dex & 0.2 & [Si/H] measurement  \\
Ni\_H\_    & float  & dex & 0.24 & [Ni/H] measurement  \\
V\_H\_    & float  & dex & 0.2 & [V/H] measurement  \\
Ca\_H\_   & float  & dex & 0.23 & [Ca/H] measurement \\
Ti\_H\_    & float  & dex & 0.19 & [Ti/H] measurement  \\
Al\_H\_    & float  & dex & 0.24 & [Al/H] measurement  \\
Y\_H\_    & float  & dex & 0.2 & [Y/H] measurement \\
prov & string  & \nodata     & brewer & catalog author \\
\sidehead{\textit{Gaia RVS Parameters:}} 
snr & integer  & \nodata    & 1301     & s/n of star's RVS spectrum\\
teff\_gspspec    & integer   & K      & 5730  & measured \teff\ from Gaia RVS \\
logg\_gspspec      & float   & dex &  4.06 & measured \logg\ from Gaia RVS \\
feh\_gspspec      & float   & dex & 0.09 & measured \feh\ from Gaia RVS \\
%
\enddata
\tablecomments{The complete table is available in the online journal.}
\end{deluxetable*}

\subsection{Gaia RVS Spectra} \label{sec:rvs_testset}
Launched in 2013, the Gaia space telescope has created the largest and most precise 3D kinematic catalog of objects in the sky. Gaia is equipped with an astrometry instrument (ASTRO), which measures positions and the photometric instrument (BP and RP filters), which measures fluxes. Its third instrument, the radial velocity spectrometer (RVS), measures line of sight velocities and is expected to obtain $ > 150$ million spectra in the Calcium (Ca) triplet region (845--870 nm) at a resolution of R=11,200 \citep{recio-blanco}. With repeat observations, the S/N will improve as these are combined \citep{dr2spec}. The third and latest data release, DR3 \citep{GaiaDR3}, provides full astrometric solutions for over 1.46 billion stars in the Galaxy. DR3 also includes spectra for just under one million stars from RVS and stellar parameters from RVS spectra including \teff, \logg, \feh, and \alphafe\ and abundances for 13 elements: N, Mg, Si, S, Ca, Ti, Cr, FeI, FeII,
Ni, Zr, Ce and Nd \citep{RecioBlanco23}. 

We are ultimately interested in finding solar analogs in the Gaia RVS spectra sample. Rather than examining all 999,645 spectra, we choose a 3D box in \teff, \logg, and \feh\ space centered around the mean Gaia-RVS reported \teff, \logg, and \feh\ of the reference set (see Table \ref{t:rs}). This way the values are homogenized in Gaia across the catalogs from the literature. We select the box to extend $3-5\sigma$ from the mean of the values for the reference set, which includes all of the reference stars. The box is defined as such: 
\teff: $5804 \pm 868$ K, \logg: $4.085\pm 1.27$ dex, and \feh: $-0.11 \pm 0.57$ dex
and contains 75,545 stars. We do a $\chi^2$ comparison with the median solar analog RVS spectrum constructed from the reference set and all 75,545 stars from the 3D box in parameter \space. Using the same $\chi^2$ thresholds when updating the reference set leaves 64,557 stars with Gaia RVS spectra that will make up our eventual test set.

We also remove the first and last wavelength steps (4 nm) from each RVS spectrum as there are often spurious flux values in these regions. This makes up 3\% of the spectrum and does not affect subsequent analyses.

\subsection{Planet host Catalogs}\label{sec:ph_cat}
\begin{figure}
    \centering
    \includegraphics[width=0.45\textwidth]{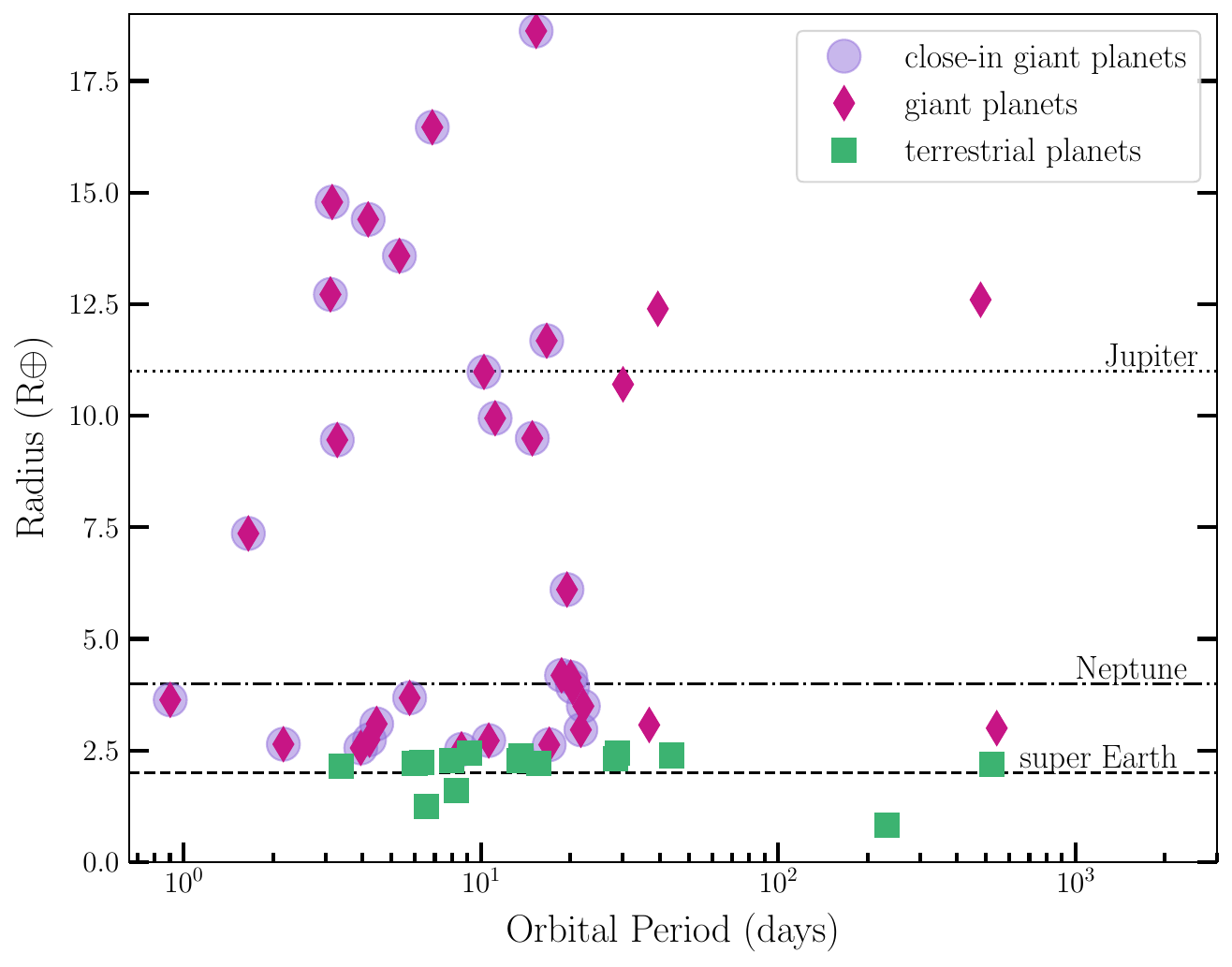}
    \caption{Distribution of planets hosted by post-analysis Gaia RVS test set stars in orbital period-radius space. Super Earth, Neptune, and Jupiter radii demarcated by dashed, dashed-dotted, and dotted lines respectively. We classify planets as close-in giant planets ($> 2.5$ R$\oplus$ , $< 30$ day orbital period) shown as purple circles, giant planet hosts ($> 2.5$ R$\oplus$) shown as pink diamonds, and ``terrestrial"/small planets ($< 2.5$ R$\oplus$) as green squares.}
    \label{fig:p-params}
\end{figure}

Our interest in solar analog planet hosts arises from the the Sun and its status as a planet host. We start by crossmatching the 64,557 Gaia RVS stars to the 7,993 planet (candidate) hosts and corresponding stellar and planet parameters from \cite{Berger23} by Gaia DR3 ID. This catalog is more expansive than the exoplanet archive as it includes $\sim 4000$ planet candidates in addition to $\sim 5000$ confirmed planets. It reports homogenized and precise stellar and planet parameters that were derived using Gaia DR3 photometry, parallaxes, spectrophotometric metallicities, and isochrone fitting \citep{Huber17,Berger20}. We follow this with a crossmatch by Gaia ID to the Exoplanet Archive \citep{ps} to catch any planet systems not in the \cite{Berger23} catalog. In Figure \ref{fig:p-params}, we show the 61 planets hosted by 50 of our test set stars in orbital period-radius space that we categorize as close-in giant planets, giant planets, and terrestrial/small planets in Section \ref{sec:phc}.


\section{Methods} \label{sec:methods}

\subsection{Applying the Cannon}\label{sec:cannon}

We use \cannon\ \citep{cannon} implementation from \cite{Ho17} to infer increased-precision stellar parameters and abundances for solar analogs identified with Gaia RVS spectra. \cannon\ is a data-driven method that can cross-calibrate between surveys and has been used successfully many times across various surveys (e.g. \citealt{Casey2017,Wheeler2020,Wylie2021,Walsen23}). \cannon\ builds a model to describe the variability in spectra using a set of reference stars (called the reference set) with known and precise stellar parameters and abundances such as \teff, \logg, \feh, \alphafe, and [X/H], which we more generally call ``labels'' following machine learning terminology. The model is then applied to the test set of stars with unknown or imprecise existing labels in the same survey. 

A linear\footnote{Because of the restricted parameter space, we find a linear model is sufficient to describe the data (which is often the case, e.g. \citealt{Hogg19,Birky20}) and that the polynomial model overfits the data.} or more often a polynomial model has been typically used with \cannon\ to connect the reference stars' labels to their spectra in the training step. If, for example, we chose a linear model made up of four labels \teff, \logg, \feh, and \alphafe\ for star $n$: each flux value for this star, $F_n$, in the star's spectrum with a number of wavelength steps, $\lambda$, would be represented as the following equation for training the model, for a set of $n$ reference objects with wavelength steps $\lambda$:
\begin{equation}
  \label{eqn:mod}
  \begin{split}
    \text{F}_{n_{\lambda}} = A_{\lambda}(\teff_n) + B_{\lambda}(\logg_n) + C_{\lambda}(\feh_n) \\ + D_{\lambda}(\alphafe_n) + E_{\lambda} + \sigma_{\lambda}.
  \end{split}
\end{equation}

 $\teff_n$, $\logg_n$, $\feh_n$, and $\alphafe_n$ are the labels for the $n$th star in the reference set vector, more generally represented as the label matrix, $l_{n}$. $A_{\lambda}$, $B_{\lambda}$, $C_{\lambda}$, $D_{\lambda}$, and $E_{\lambda}$ are the coefficients for which \cannon\ solves and are more generally represented by the coefficient matrix, $\theta_\lambda$. $\sigma_{\lambda}$ is the noise term and is the root-mean-squared (rms) sum of the observational error on each flux value ($\sigma_{n,\lambda}$) and the scatter in the model at each wavelength ($s_\lambda$). 

During \cannon's training step, Equation \ref{eqn:mod} is re-represented as a single-wavelength step log-likelihood function that is optimized to solve for $\theta_\lambda$ and $s_{\lambda}$ using the input $l_{n}$ and the input flux and uncertainty for the $n$ reference objects. Then, to infer the label matrix for the test set of $m$ stars, $l_{m}$, the same log-likelihood is optimized, but $\theta_\lambda$ and $s_\lambda$ are held constant instead. Additionally, the optimization is done as a least squares fit over the whole spectrum rather than per wavelength step. 

We build a reference set of 85 solar analogs' labels measured in high-resolution (R $ > 30,000$) spectroscopic surveys as described in Section \ref{sec:lsa}. We identify the reference stars associated with high S/N Gaia RVS spectra and train a 17-label linear model (\teff, \logg, \feh [Mg/H], [C/H], [N/H], [O/H], [Na/H], [Mn/H],[Cr/H], [Si/H], [Ni/H], [V/H], [Ca/H], [Ti/H], [Al/H], [Y/H]) on the high-precision labels and Gaia RVS spectra with \cannon. We then use this model to infer labels for 64,557 spectroscopically similar Gaia RVS stars (as determined and described in Section \ref{sec:rvs_testset}). 

\begin{figure*}
    \centering
    \includegraphics[width=\textwidth]{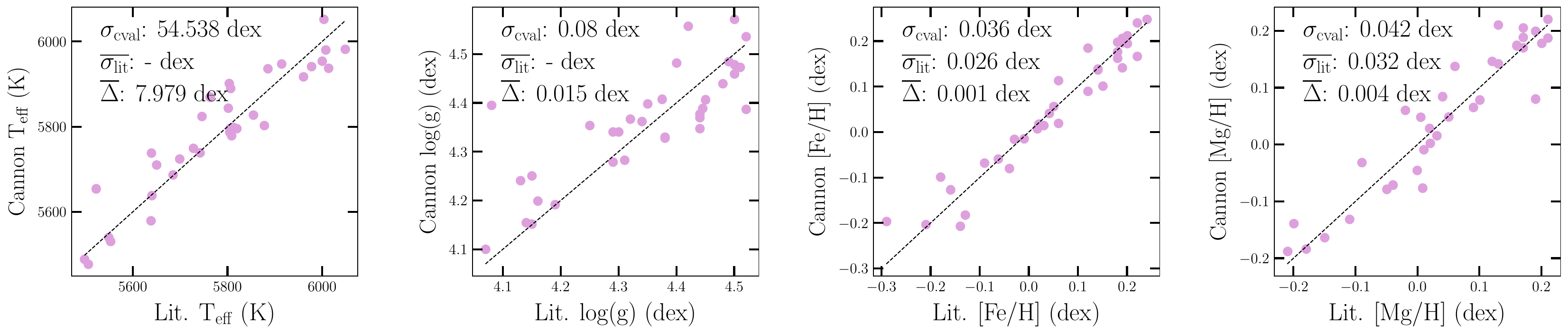}
    \includegraphics[width=\textwidth]{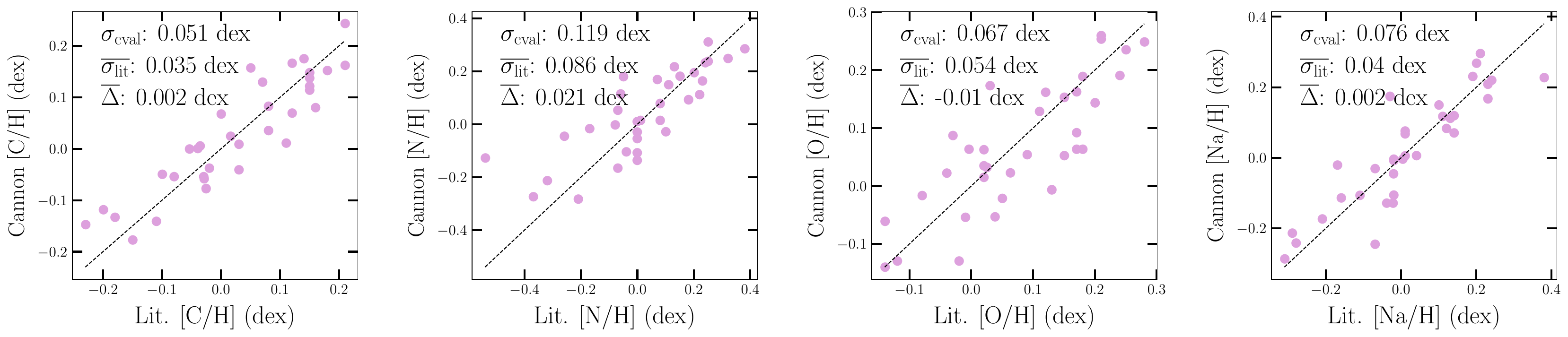}
    \includegraphics[width=\textwidth]{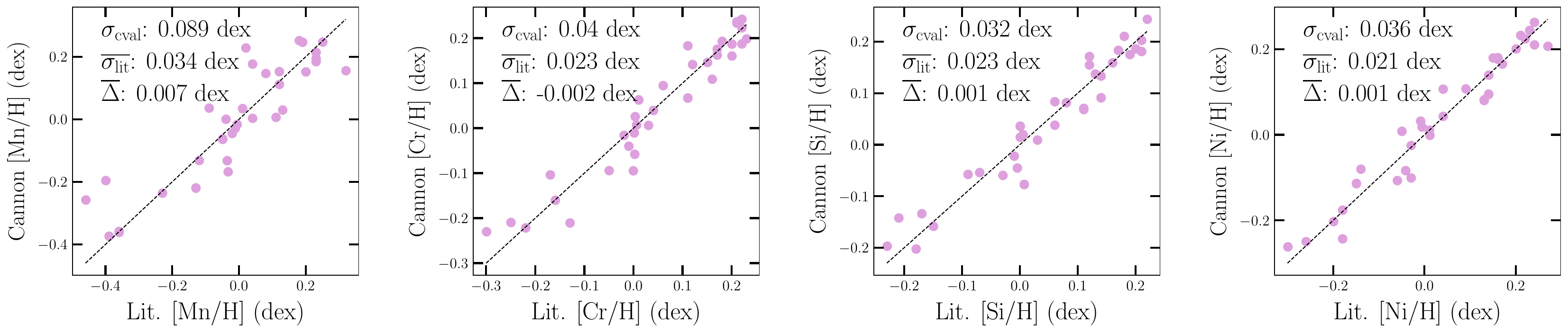}
    \includegraphics[width=\textwidth]{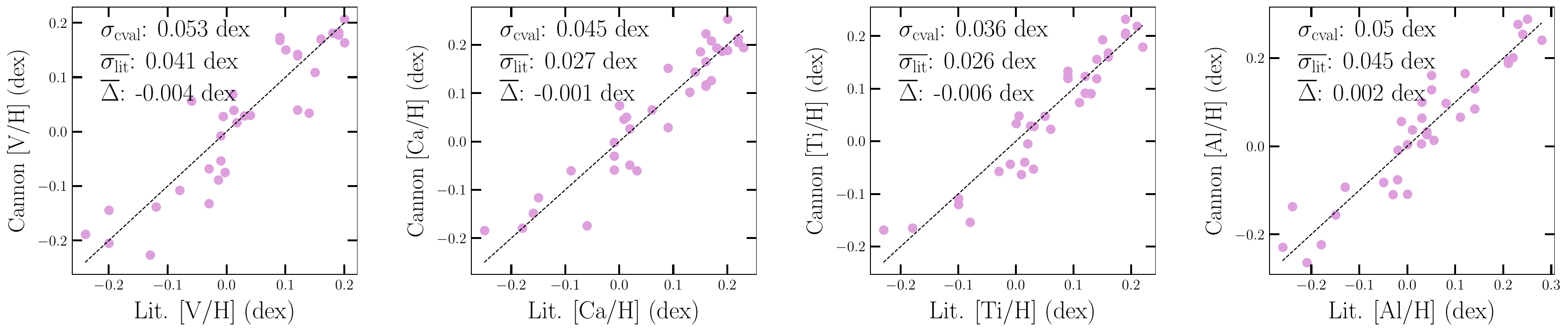}
    \includegraphics[width=\textwidth]{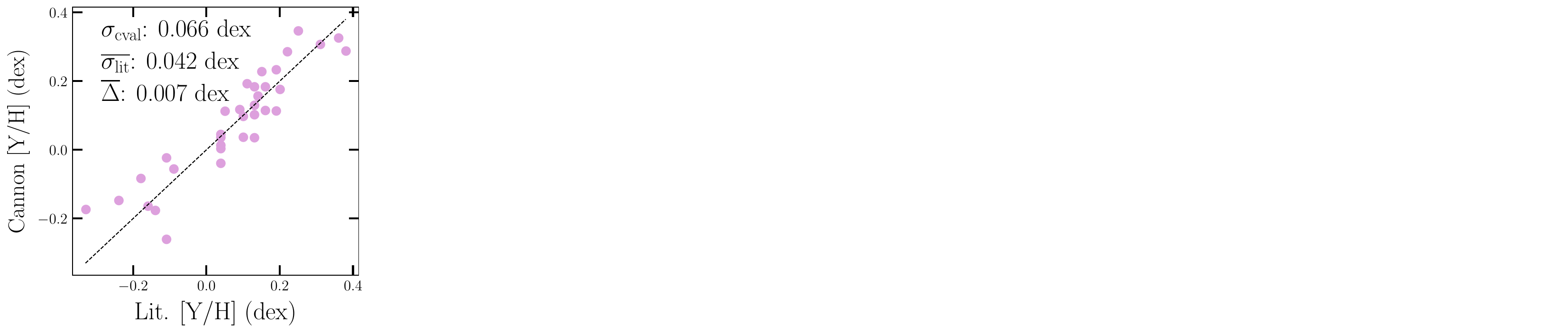}

    \caption{Literature label values versus \cannon-model predicted label values with 1:1 ratio as black dashed line, the $1\sigma$ standard deviation of predicted values from \cannon\ -- literature values ($\sigma_{\rm cval}$), the mean error on the label from the literature ($\overline{\sigma_{\rm lit}}$), and the bias (or the mean difference in predicted and literature values, $\overline{\Delta}$). $\overline{\sigma_{\rm lit}}$ was not available for \teff\ and \logg\ as these were not reported for stars from \cite{Hinkel14}.}
    \label{fig:crossval_scatters}
\end{figure*}

\begin{figure*}
    \centering
    \includegraphics[width=\textwidth]{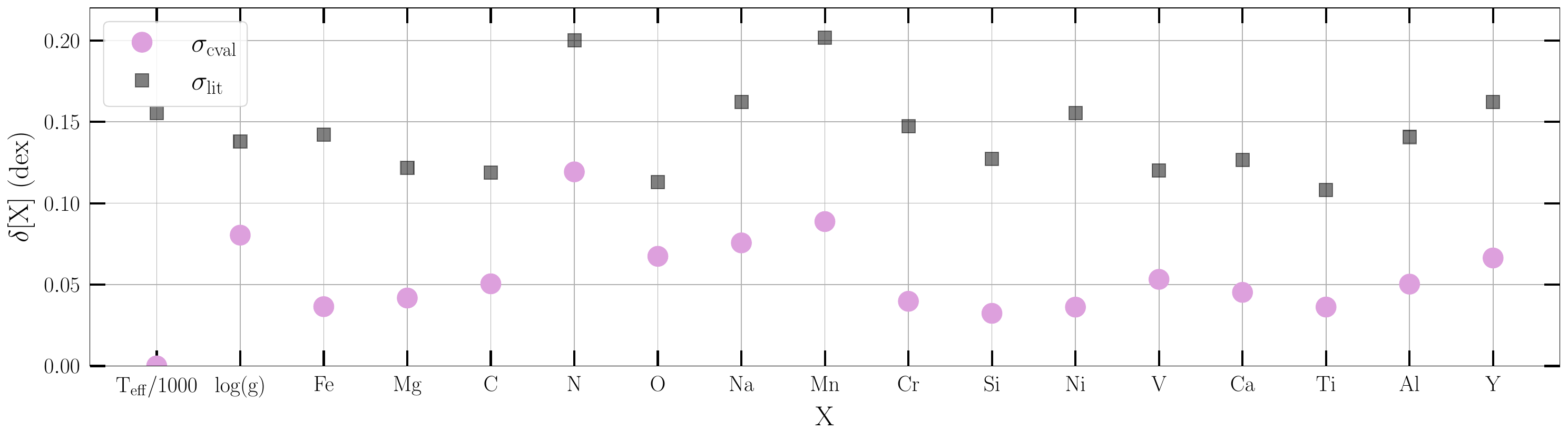}
    \caption{Summary of cross-validation results for each label. Standard deviation (1$\sigma$) of predicted values from \cannon\ -- literature values ($\sigma_{\rm cval}$) are plotted as pink circles. Standard deviation (1$\sigma$) of reference set's literature values ($\sigma_{\rm int}$) are plotted as gray squares.}
    \label{fig:crossval_sum}
\end{figure*}

\begin{figure*}
    \centering
    \includegraphics[width=\textwidth]{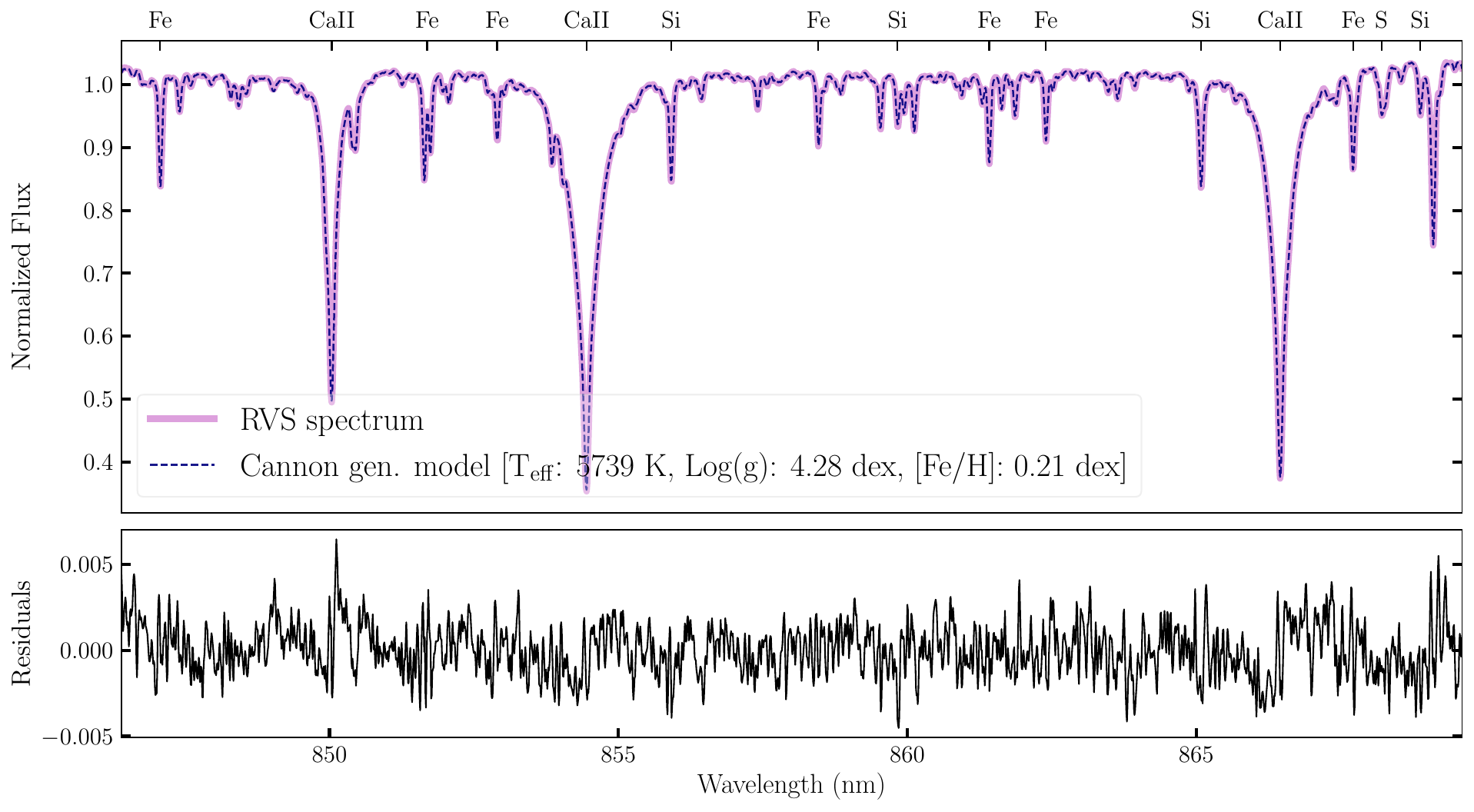}
    \caption{\textit{Top}: Example RVS spectrum from reference set (pink) with generated model spectrum (blue dashed line) from \cannon\ cross-validation and inferred labels of \teff: 5739 K, \logg: 4.28 dex, and \feh: 0.21 dex. \textit{Bottom}: Residuals of data -- model in black, which are on the order of 0.5\%.}
    \label{fig:genmodel_ex}
\end{figure*}

\subsubsection{Testing Validity of Reference Set}\label{sec:refset}
In order to apply \cannon\ to our test set of $> 60,000$ stars, it is important to ascertain that the reference set and associated model are robust. We employ a cross-validation test after training time and prior to test time, where we iteratively treat each star in the reference set as a ``test star''. For a given test star, we remove its spectrum and labels from the reference set, and \cannon\ trains on the rest of the reference set to predict the test star's labels and spectrum. We run this procedure for every star. In turn, we are able to evaluate, for each label in the reference set (\teff, \logg, \feh, [Mg/H], [C/H], [N/H], [O/H], [Na/H], [Mn/H], [Cr/H], [Si/H], [Ni/H], [V/H], [Ca/H], [Ti/H], [Al/H], and [Y/H]), how well the model's inferred values match the reference set values from the literature.

For a single label, if the $1\sigma$ standard deviation of the ensemble's \cannon-predicted values -- the literature values, or ($\sigma_{\rm cval}$) is less than the 1$\sigma$ standard deviation across the literature values in the reference set ($\sigma_{\rm lit}$), this shows the model learned this particular label. If $\sigma_{\rm cval} \geq \sigma_{\rm lit}$, this means the model is doing no better than randomly drawing from the reference label distribution. For example, for the \feh\ label, the 1$\sigma$ standard deviation in the literature values is 0.14 dex. \cannon\ can predict the literature \feh\ to an average precision of $\sigma_{\rm cval} =$ 0.04 dex, which is much smaller than the 1$\sigma$ standard deviation of the reference set \feh\ values indicating \cannon\ model is learning the label of \feh\ from the spectra. 

\cannon\ requires a reference set with a high S/N. We find that an S/N $\geq 50$ (45 stars) is sufficient in that the standard deviation of the predicted--literature values is generally less than the standard deviation of the literature values for each label. We find improved precision with an S/N $\geq 100$ (34 stars) and adopt this limit. With any higher S/N, we have too few stars to build a reliable model because the number of stars reaches the number of coefficients in the model. The cross-validation test with these 34 stars results in a standard deviation of 55 K in \teff, 0.08 dex in \logg, 0.036 dex in \feh, 0.042 in [Mg/H], 0.051 dex in [C/H], 0.119 dex in [N/H], 0.067 dex in [O/H], 0.076 dex in [Na/H], 0.089 dex in [Mn/H], 0.04 dex in [Cr/H], 0.032 dex in [Si/H], 0.036 dex in [Ni/H], 0.053 dex in [V/H], 0.045 dex in [Ca/H], 0.036 dex in [Ti/H], 0.05 dex in [Al/H], and 0.066 dex in [Y/H]. We show the literature values versus \cannon\ values with a 1:1 line, the 1$\sigma$ standard deviation of predicted values from \cannon\ -- the literature values ($\sigma_{\rm cval}$), the mean error on the label from the literature ($\overline{\sigma_{\rm lit}}$), and the bias (or the mean difference in predicted and literature values, $\overline{\Delta}$) in Figure \ref{fig:crossval_scatters}.

We see [N/H] has the highest standard deviation, and that there is a pile-up of stars at a single literature value of [N/H] in Figure \ref{fig:crossval_scatters} (second row and second panel). These are the stars from \cite{bedell18} that did not have reported abundances for N. As discussed in Section \ref{sec:lsa}, for reference stars that did not have certain labels, we assign them the mean of the population as \cannon\ does not formally handle partial labels. This means that these stars are ``badly labeled", which in turn yields poorer cross-validation results (e.g. a higher standard deviation) for N. In previous iterations, we included the stars from \cite{Bensby14} which did not have measurements for 4 abundances and also made up 1/4 of the reference set. This yielded poor cross-validation results that also affected subsequent analyses, so we exclude them from our reference set when using \cannon. 


 We compare $\sigma_{\rm cval}$ (as pink circles) with $\sigma_{\rm lit}$ (as gray squares) for each label; this is summarized in Figure \ref{fig:crossval_sum}. No $\sigma_{\rm cval}$ for any label exceeds $\sigma_{\rm lit}$. With the newly inferred labels, \cannon\ can also generate a model spectra to compare to the observed spectra. We show an example of \cannon-generated model spectrum in the blue dashed line and the RVS spectrum in pink for a reference star in the top panel of Figure \ref{fig:genmodel_ex}. The residuals are plotted in black in the bottom panel and are on the scale of 0.5\% or less, indicating the model is working very well. The maximum of each reference star's residuals found in the reference set are on average $1.5 \pm 0.9\%$.

\subsubsection{Inferring Labels for Test Set}\label{sec:testset_methods}
Given the robust performance of the model during cross-validation, we train the model on the entire high S/N reference set (34 stars) and apply 
\cannon\ model to our test set of 64,557 stars. 
For each set of labels for each star, we have an associated error that is the rms sum of $\sigma_{\rm cval}$ and the unique noise term, $\sigma_{\lambda}$, returned in the star's covariance matrix. We use the labels and associated errors to generate a model spectrum and calculate a $\chi^{2}$ goodness of fit for each star. 
We remove any stars with a $\chi^{2} > 4642$, which is two times the number of degrees of freedom, or the number of wavelength steps in the spectrum ($\lambda = 2321$). We are left with 63,725 stars. \cannon\ can extrapolate in label inference beyond the domain of values set by the range of the reference set labels. To keep our results robust, we remove anything outside of the reference set labels and are left with 17,412 stars. This drop is mainly attributed to stars being too hot or too cold compared to the temperature range of the reference set. As mentioned in Section \ref{sec:ph_cat}, 43 of these stars have been identified as planet (candidate) hosts.

\subsection{Inferring Isochrone Ages from Test Labels}
Since the reference set does not have precise reported ages, we do not include age in our model. However, we are interested in exploring the effects of age in our work. We recognize that because these are main-sequence field stars stars, the ages inferred using stellar models will be imprecise. We fit isochrones to each star and coarsely estimate ages that we use as points of exploration. We do not report these as robust results for individual stars. However they serve as a measure of the overall population age distribution.

We infer ages for our test set with \texttt{isoclassify} \citep{Huber17,Berger20}. \texttt{isoclassify} uses an isochrone grid modeling approach. Isochrones from the MIST database \citep{Dotter16,Choi16,Paxton11,Paxton13,Paxton15} are interpolated and used to calculate a grid of ages from 0.5--14 Gyr (step size of 0.25 Gyr) and metallicities from -2--0.4 dex (step size 0.02 dex). Given observables, \texttt{isoclassify} calculates priors and likelihoods and integrates over all points in the grid to derive posterior distributions following \cite{Serenelli13}.

\begin{figure}
    \centering
    \includegraphics[width=0.5\textwidth]{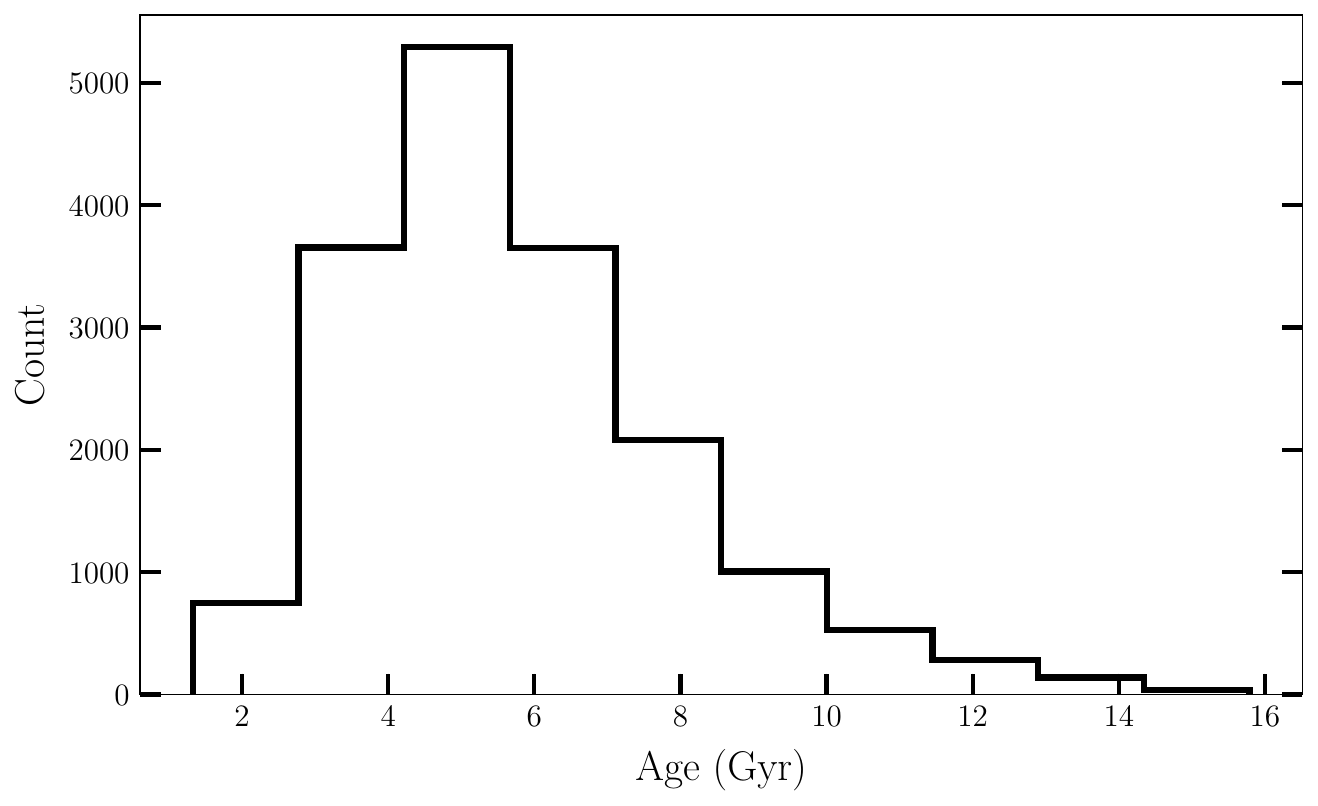}
    \caption{Approximate age distribution of test set using \teff, \logg, and \feh\ as inputs in \texttt{isoclassify}. Stars are on average $5.8\pm2.3$ ($1\sigma$)~Gyr.}
    \label{fig:ages}
\end{figure}

For simplicity, we only use the spectroscopic properties, \teff, \logg, and \feh\ inferred from \cannon, as these are intrinsic properties to the star. We do not account for reddening as 90\% of our stars have an A$_\text{V}$ $< 0.2$ from the calculated reddening in the Gaia bandpasses using \texttt{Bayestar19} \citep{Green3D2019}, so these properties inferred from the spectra should not be affected. We find that these stars are on average $5.8\pm2.3$ Gyr and show the distribution of ages in Figure \ref{fig:ages}. The average age error across the test set is $1.84 \pm 0.77$ Gyr.

\begin{deluxetable*}{lcccl}
\tablecaption{Test Set Inferred Labels \label{t:ts}}
\tablewidth{0pt}
\tablehead{
\colhead{Column} & \colhead{Format} & \colhead{Units} & \colhead{Example} & \colhead{Description}
}
\startdata
\sidehead{\textit{Identifier:}} 
DR3id  & integer & \nodata & 711643129958302592 & Gaia DR3 ID \\
\sidehead{\textit{Coordinates:}} 
ra      & float & degrees & 134.765818 & Right ascension \\
dec     & float & degrees & 30.568571 & Declination \\
\sidehead{\textit{Cannon Inferred Labels:}} 
teff   & integer  & K    & 5915 & \teff\ inferred label \\
teff\_err   & integer  & K    & 55 & \teff\ inferred label error  \\
logg   & float  & dex    & 4.31 & \logg\ inferred label  \\
logg\_err   & float  & dex    & 0.08 & \logg\ inferred label error  \\
Fe\_H\_     & float  & dex    & -0.01 & \feh\ inferred label  \\
Fe\_H\_err     & float  & dex    & 0.03 & \feh\ inferred label error  \\
Mg\_H\_   & float  & dex    & -0.02 & [Mg/H] inferred label \\
Mg\_H\_err    & float  & dex    & 0.04  & [Mg/H] inferred label error \\
C\_H\_     & float  & dex     & -0.008 & [C/H] inferred label \\
C\_H\_err     & float  & dex     & 0.05 & [C/H] inferred label error \\
N\_H\_    & float  & dex     & 0.007 & [N/H] inferred label \\
N\_H\_err    & float  & dex     & 0.12 & [N/H] inferred label error \\
O\_H\_   & float  & dex & 0.05& [O/H] inferred label \\
O\_H\_err   & float  & dex & 0.07 & [O/H] inferred label error \\
Na\_H\_    & float  & dex & -0.07 & [Na/H] inferred label\\
Na\_H\_err    & float  & dex & 0.08 & [Na/H] inferred label error\\ 
Mn\_H\_    & float  & dex & -0.1 & [Mn/H] inferred label \\
Mn\_H\_err    & float  & dex & 0.09 & [Mn/H] inferred label error \\
Cr\_H\_    & float  & dex & -0.03 & [Cr/H] inferred label  \\
Cr\_H\_err    & float  & dex & 0.04 & [Cr/H] inferred label error  \\
Si\_H\_    & float  & dex &-0.005 & [Si/H] inferred label  \\
Si\_H\_err    & float  & dex & 0.03 & [Si/H] inferred label error  \\
Ni\_H\_    & float  & dex & -0.04 & [Ni/H] inferred label  \\
Ni\_H\_err    & float  & dex & 0.03 & [Ni/H] inferred label error  \\
V\_H\_    & float  & dex & -0.05 & [V/H] inferred label  \\
V\_H\_err    & float  & dex & 0.05 & [V/H] inferred label error  \\
Ca\_H\_   & float  & dex & 0.005 & [Ca/H] inferred label \\
Ca\_H\_err   & float  & dex & 0.05 & [Ca/H] inferred label error \\
Ti\_H\_    & float  & dex & 0.008 & [Ti/H] inferred label  \\
Ti\_H\_err    & float  & dex & 0.04 & [Ti/H] inferred label error \\
Al\_H\_    & float  & dex & -0.05 & [Al/H] inferred label  \\
Al\_H\_err    & float  & dex & 0.05 & [Al/H] inferred label error  \\
Y\_H\_    & float  & dex & 0.02 & [Y/H] inferred label \\
Y\_H\_err    & float  & dex & 0.07 & [Y/H] inferred label error \\
\sidehead{\textit{Other:}} 
age    & float   & Gyr      & 5.84  & derived age from \texttt{isoclassify}\\
age\_err      & float   & Gyr &  2.67 & derived error in age from \texttt{isoclassify}\\
planet? & integer  & \nodata    & 0     & does star host planet(s)? N=0, Y=1 
%
\enddata
\tablecomments{The complete table is available in the online journal. Values have been truncated in this table preview to remain concise. [N/H] was inferred using partial labels in the training step.}
\end{deluxetable*}

\section{Results}\label{sec:results}
We successfully infer labels for 17,412 stars. These stars have inferred labels that are within the label range of our reference set and a reasonable $\chi^{2}$ fit (discussed in Section \ref{sec:testset_methods}) between the RVS spectra and \cannon-generated models. We include a table of the 17 labels and label errors inferred with \cannon, age and age errors from \texttt{isoclassify}, and planet host status in Table \ref{t:ts}. The average error value on each label is $61$ K for \teff, $0.09$ dex for \logg, and $0.04-0.1$ dex for [X/H].

\begin{figure*}
    \centering
    \includegraphics[width=\textwidth]{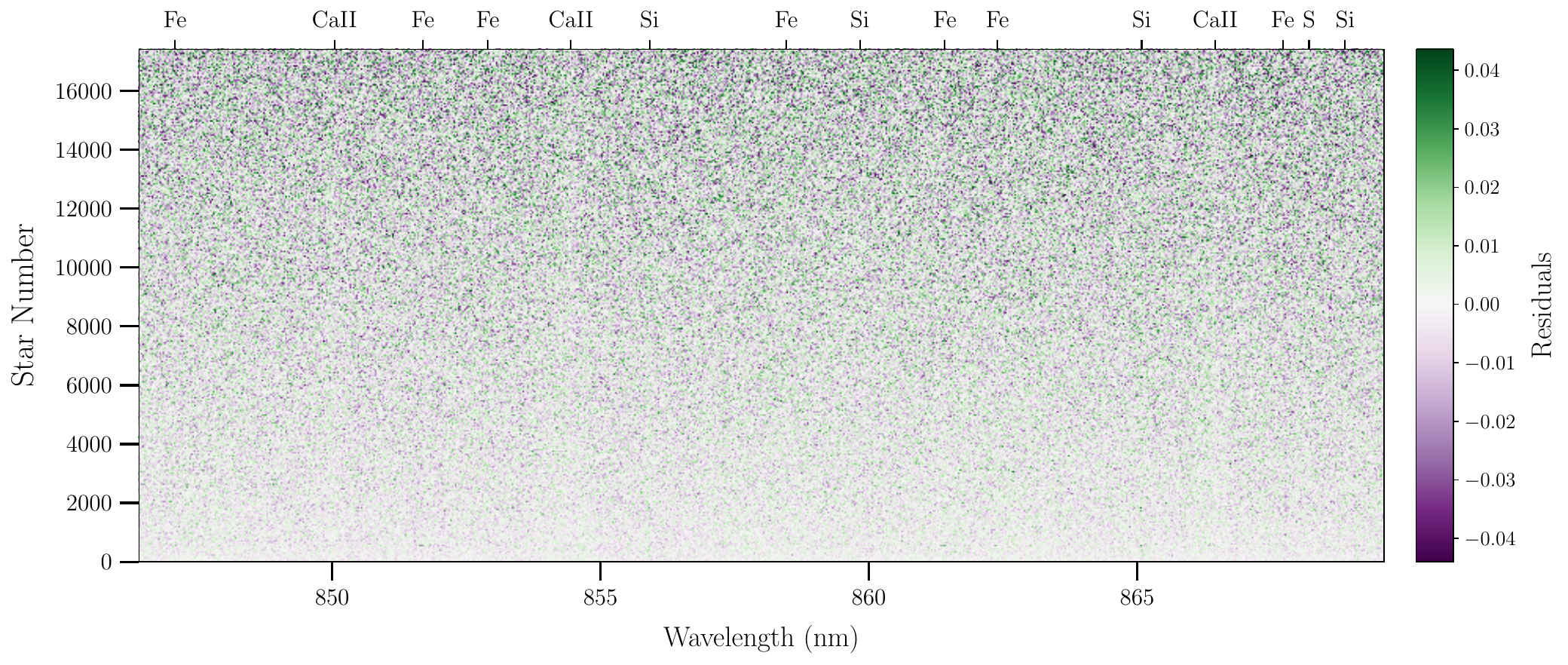}
    \caption{Residuals of test set ordered from lowest (bottom) to highest S/N (top). The Ca triplet is remarkably well modeled. All other residuals appear stochastic in nature.}
    \label{fig:imshow}
\end{figure*}

We subtract the model spectra generated with the stars' labels from the observed RVS spectra and show the residual flux vectors of our test set in Figure \ref{fig:imshow} arranged from lowest (bottom) to highest S/N (top). The residuals look fairly stochastic with marginally higher residuals for high S/N stars but still typically sub-percent amplitudes. On average, we find residuals of $0.04 \pm 0.03\%$. 

Of note are the Ca triplet lines, which have residuals of near zero indicating that the model performs very well. This is a slight departure from \cite{Rampalli21}, where the strongest residuals were seen at the Ca lines. The cores of the Ca lines are formed in the stellar chromosphere \citep[e.g.][]{Andretta2005}. These cannot be properly modeled under the typical one dimensional, Local Thermal Equilibrium (1D LTE) assumption invoked when deriving stellar parameters \citep{Magic2013}. An important difference between our prior work and this work is that here we restrict our study to a much smaller parameter space that describes solar analogs, and we see that the linear model can well describe the flux given the labels. In \cite{Rampalli21}, we covered a broader range of \logg\ from 2 to 3 using a second order polynomial model to describe red clump stellar spectra. The Ca line residuals found there were already small at the percentage level. Therefore, it is not surprising these residuals decrease significantly in our work and are not necessarily at odds with previous results. 

We show the coefficients of the model for each label in Figure \ref{fig:coeffs} of Appendix \ref{sec:modelcoeffs}. This shows at what wavelengths the model is learning from for each label. We see that most have the highest amplitudes at the cores and wings of the Ca lines similar to \cite{Rampalli21}. While there are shared regions of the spectra that the model learns from for each label, the coefficient shapes are in detail different. 

    


\subsection{The Dimensionality and Precision Limits of Gaia-RVS Spectra}
In large-scale and low to medium resolution spectroscopic surveys, the concept of dimensionality has been explored at great length (e.g. \citealt{ting12, pricejones,Griffith21,Ting2021}): how many necessary labels are needed to describe the variance seen in the spectra and the chemical abundance measurements derived from these spectra? We explore the dimensionality of Gaia-RVS spectra by seeing how well the four fundamental labels, \teff, \logg, \feh, and \mg\ ([Mg/H]--\feh\ representing alpha enrichment, \alphafe) inferred from \cannon\ and ages from \texttt{isoclassify} can predict the other 13 abundances inferred from \cannon. If we can perfectly predict the 13 abundances within their errors from these 5 labels, this indicates the data are low dimensional and that the labels are all entirely correlated. This means that we are not gleaning any new information from the elemental abundances that is not already reported in the \teff, \logg, \feh, \mg, and age measurements. The measurement precision that can be reached by a spectroscopic survey increases with instrument resolution. Prior work in this realm indicates that very high resolution, high fidelity data (e.g. from the HARPS spectrograph, \citealt{HARPS}) is required to access the information in stellar spectra beyond that captured in a few labels (e.g. \citealt{Ness22}).  

Following the methodology outlined in Section \ref{sec:cannon}, we implement a linear regression with Python's \texttt{sklearn} to predict element abundances using the five fundamental labels of \teff, \logg, [Fe/H], \mg, and age. Equation \ref{eqn:mod} is invoked again but instead of having the labels and coefficients describe a vector of flux values, the vector is substituted with a single abundance value. We find that all 13 element labels from \cannon\ are predicted with the 5 label regression to within errors. This is consistent with results using other similar resolution surveys such as ARGOS and RAVE \citep{Rampalli21,Casey2017}. \cite{Griffith21} show that even with higher resolution surveys like APOGEE (R=22,500), a subset of elements, \mg\ and [Mg/H], is all that is needed to predict other abundances. We test this by using a linear regression model with only \mg\ and [Mg/H] and subsequently \feh\ and \mg\ and find that in both cases indeed 2 labels is enough to predict the other abundances within their errors. 

It is important to note that the errors reported here, though dominated by  the cross validation term, $\sigma_{\rm cval}$, are upper limit errors. This is because the error in the reference labels are folded into these numbers \citep{Manea23}. We can measure the intrinsic dispersion, $\sigma_{\rm int}$ to better assess \cannon's performance and Gaia RVS. $\sigma_{\rm int}$ is the difference of $\sigma_{\rm cval}$ and the mean error ($\sigma_{\rm int}$) in quadrature for each label. We calculate a $\sigma_{\rm int}$ of 0.02 for Si and Al all the way to 0.08 dex for N due to bad input labels as discussed in Section \ref{sec:refset}. We report $\sigma_{\rm int}$ for all of the abundances in Table \ref{t:intdisps}. If we consider the 2 label model regression results using intrinsic dispersions rather than our reported errors, we still find that we can 100\% predict 7/13 abundances. There are 6 exceptions: C (97\%), N (94\%), O (87\%), Ti (98\%), Al (92\%), and Y (98\%).  
\begin{deluxetable}{cccc}[h!]
\centering 
\tabletypesize{\footnotesize} 
\tablecolumns{4}
\tablewidth{0pt}
\tablecaption{$\sigma_{\rm int}$ for Abundance Measurements}
\label{t:intdisps}
\tablehead{
\colhead{Element} & 
\colhead{$\sigma_{\rm cval}$ (dex)} & 
\colhead{$\overline{\sigma_{ \rm lit}}$ (dex)} &
\colhead{$\sigma_{\rm int}$ (dex)}
}
\startdata
Fe & 0.036 & 0.026 & 0.025 \\
Mg & 0.042  & 0.032  & 0.026  \\
C & 0.051 & 0.035 & 0.037 \\
N & 0.119 & 0.086 & 0.082 \\
O & 0.067 & 0.054 & 0.041 \\
Na & 0.076 & 0.04 & 0.064 \\
Mn & 0.089 & 0.034 & 0.082 \\
Cr & 0.04 & 0.023 & 0.033 \\
Si & 0.032 & 0.023 & 0.023 \\
Ni & 0.036 & 0.021 & 0.03 \\
V & 0.053 & 0.041 & 0.034 \\
Ca & 0.045 & 0.027 & 0.036 \\
Ti & 0.036 & 0.026 & 0.025 \\
Al & 0.05 & 0.045 & 0.023 \\
Y & 0.066 & 0.042 & 0.051 \\
\enddata

\end{deluxetable}

Thus, even when considering reference label errors, we still find the abundance data are low dimensional, consistent with previous studies \citep{Casey2017,Rampalli21,Griffith21}. Obtaining intrinsic abundance information, beyond the 2--5 labels, can only be accessed if the abundance precision is higher than the intrinsic dispersion measurements for these abundances ($ < 0.02$ dex) as also found in \cite{Behmard23}. In spite of the low precision on individual measurements, we are able to test for potential planet host differences in the amplitudes of the element abundances, in particular with condensation temperature, given the size of our data set, as demonstrated in the next section. Because the planet sample is a small portion of the test set, differences in individual abundances as a function of planet status could be present at a level lower than our overall precision. 

\subsection{Planet host Comparisons}\label{sec:phc}
Of the 17,412 stars in our test set, 50 are known planet (candidate) hosts as shown in Figure \ref{fig:p-params}. The 61 planets span a range of radii and orbital periods. We list the planet parameters and host star information in Table \ref{t:phs}. For our analysis, we further categorize the sample of planet hosts into the 27 stars hosting close-in giant planets (R$\oplus$ $> 2.4$, orbital period $< 30$ days), the 35 stars hosting giant planets ($> 2.5$ R$\oplus$), and the 15 stars that lack known close-in or distant giant planets and host ``terrestrial" or small planets ($< 2.5$ R$\oplus$). The R $=$2.5R$\oplus$ boundary is chosen to have the closest radius value to terrestrial-sized planet radii (typically $\leq 2$ R$\oplus$) that still maintains a statistically substantial population of stars. We recognize that the delineation between “terrestrial” planets and giant planets is somewhat arbitrarily drawn. We conduct the following comparison tests using a cut of 3, 4, and 5R$\oplus$ and ultimately find the same results. We also call stars in the test set without identified planets \textit{non-planet hosts} though we recognize it is highly likely that many of these stars host as-yet-undetected planets since 50--70\% of Sun-like stars host planets \citep{Mulders18,Zink19}. 

\startlongtable
\begin{deluxetable*}{ccccccc}
\centering 
\tabletypesize{\footnotesize} 
\tablecolumns{4}
\tablewidth{0pt}
\tablecaption{Planet (candidate) parameters from \cite{Berger23} and the Exoplanet Archive for planet hosts in test set.}
\label{t:phs}
\tablehead{
\colhead{DR3} & 
\colhead{host id} & 
\colhead {ra} &
\colhead{dec} & 
\colhead{planet id} & 
\colhead{radius (R$\oplus$)} &
\colhead{orbital period (days)}
}
\startdata
    749676822006222464 & tic85293053 & 156.8377 & 34.39095 & toi1772.01 & 2.325 & 8.051 \\ 
        749676822006222464 & tic85293053 & 156.8377 & 34.39095 & toi1772.02 & 4.493 & \nodata \\ 
        6258810550587404672 & HD 137496 & 231.74206 & -16.50901 & HD 137496 b & 1.31 & 1.621 \\ 
        6258810550587404672 & HD 137496 & 231.74206 & -16.50901 & HD 137496 c & 12.6 & 479.9 \\ 
        5500031730804862720 & tic350931281 & 90.71014 & -55.53292 & toi4403.01 & 3.097 & 4.462 \\ 
        4648587831980764160 & tic141412823 & 84.23719 & -75.61142 & toi4421.01 & 2.964 & 21.701 \\ 
        4716158250340258944 & HD 10180 & 24.47311 & -60.51149 & HD 10180 c & 3.68 & 5.76 \\ 
        4716158250340258944 & HD 10180 & 24.47311 & -60.51149 & HD 10180 d & 3.45 & 16.357 \\ 
        4716158250340258944 & HD 10180 & 24.47311 & -60.51149 & HD 10180 e & 5.39 & 49.748 \\ 
        4716158250340258944 & HD 10180 & 24.47311 & -60.51149 & HD 10180 f & 5.24 & 122.744 \\ 
        4716158250340258944 & HD 10180 & 24.47311 & -60.51149 & HD 10180 g & 4.91 & 604.67 \\ 
        4716158250340258944 & HD 10180 & 24.47311 & -60.51149 & HD 10180 h & 9.4 & 2205.0 \\ 
        4729324799004633472 & tic197807043 & 53.46724 & -57.62129 & toi2423.01 & 14.837 & \nodata \\ 
        4666490423895957376 & tic388106759 & 62.854 & -69.47488 & toi810.01 & 2.32 & 28.306 \\ 
        4666490423895957376 & tic388106759 & 62.854 & -69.47488 & toi810.02 & 2.414 & 90.849 \\ 
        4666498154837086208 & tic25155310 & 63.37473 & -69.22659 & toi114.01 & 9.459 & 3.289 \\ 
        6411016151376518144 & tic403135192 & 333.55961 & -59.56832 & toi2226.01 & 3.636 & 0.902 \\ 
        1433586454781584896 & tic224596152 & 254.5033 & 57.30526 & toi1734.01 & 2.435 & 28.874 \\ 
        4903786336207800576 & tic281781375 & 12.97603 & -59.34367 & toi204.01 & 2.391 & 43.828 \\ 
        4910452812646300544 & tic206541859 & 18.04853 & -56.92538 & toi4406.01 & 10.706 & 30.082 \\ 
        636363799347569408 & epic211945201 & 136.5742 & 19.40205 & epic211945201.01 & 6.106 & 19.492 \\ 
        636363799347569408 & epic211945201 & 136.5742 & 19.40205 & epic211945201.01 & 6.206 & 19.492 \\ 
        2581918597853527424 & epic220621788 & 12.77001 & 9.51677 & epic220621788.01 & 2.378 & 13.682 \\ 
        2079613685738988800 & kic8176564 & 295.43967 & 44.03901 & K02720.01 & 1.256 & 6.572 \\ 
        5710154317045164416 & tic1003831 & 130.29515 & -16.03633 & toi564.01 & 7.363 & 1.651 \\ 
        5762607889340459008 & tic149845414 & 135.5458 & -3.42082 & toi2545.01 & 2.274 & 7.994 \\ 
        3486043573401367168 & tic98957720 & 181.0008 & -28.32535 & toi3501.01 & 18.622 & 15.349 \\ 
        4822523493384074496 & tic167661160 & 79.70936 & -36.03769 & toi2479.01 & 3.073 & 36.838 \\ 
        4655671573080397056 & tic231077395 & 67.65781 & -70.35491 & toi2238.01 & 2.147 & 3.39 \\ 
        4743138925656526976 & tic201793781 & 34.68431 & -54.85972 & toi248.01 & 2.208 & 5.991 \\ 
        4762694805108531840 & tic382045742 & 79.32368 & -59.06839 & toi3355.01 & 2.215 & 15.66 \\ 
        1602005522755589760 & tic224313733 & 234.09914 & 57.89074 & toi1856.01 & 12.396 & 39.402 \\ 
        1630906044157332224 & tic307958020 & 256.8431 & 62.47554 & toi4633.01 & 3.002 & 543.878 \\ 
        5054967123443063936 & tic142868621 & 50.94274 & -31.2652 & toi2446.01 & 4.184 & 18.686 \\ 
        3604866386264961792 & epic212357477 & 202.01618 & -15.93798 & epic212357477.01 & 2.236 & 6.327 \\ 
        3616931735377523712 & K2-292 & 205.37587 & -9.94607 & K2-292 b & 2.63 & 16.984 \\ 
        6239702034929248512 & K2-287 & 233.07434 & -22.35835 & K2-287 b & 9.494 & 14.893 \\ 
        2534280057557160704 & epic220207374 & 20.992 & 0.25935 & epic220207374.01 & 1.608 & 8.268 \\ 
        5820908638718592256 & tic361711730 & 243.88501 & -69.2171 & toi5027.01 & 10.983 & 10.244 \\ 
        2104104723128736128 & kic6266866 & 284.59168 & 41.63933 & K05254.01 & 0.832 & 232.916 \\ 
        6385929006882225024 & tic405425498 & 337.51776 & -67.85014 & toi2227.01 & 2.744 & 4.222 \\ 
        4964562700427198720 & tic138727432 & 32.91429 & -37.89299 & toi853.01 & 16.464 & 6.865 \\ 
        3777506754255516800 & tic124573851 & 158.9007 & -5.18136 & toi669.01 & 2.559 & 3.945 \\ 
        3778075717162985600 & tic169226822 & 160.55877 & -3.835 & toi675.01 & 14.399 & 4.178 \\ 
        6173451477191225984 & tic271168962 & 210.19357 & -30.58361 & toi828.01 & 13.582 & 5.322 \\ 
        4813630986935633152 & tic13883872 & 73.74111 & -41.34421 & toi4322.01 & 2.267 & 13.419 \\ 
        5260585074966996864 & tic141770198 & 91.8755 & -76.69058 & toi4555.01 & 2.192 & 522.627 \\ 
        5262245367587966208 & tic142087638 & 98.44214 & -74.1901 & toi2404.01 & 3.924 & 20.363 \\ 
        5262245367587966208 & tic142087638 & 98.44214 & -74.1901 & toi2404.02 & 8.978 & 74.605 \\ 
        5262245367587966208 & tic142087638 & 98.44214 & -74.1901 & toi2404.03 & 6.941 & 746.066 \\ 
        3406687485600728192 & epic247098361 & 73.7668 & 18.65432 & epic247098361.01 & 9.945 & 11.169 \\ 
        6266769537305193856 & epic250106132 & 236.44605 & -14.20768 & epic250106132.01 & 3.496 & 22.122 \\ 
        6283723285046532864 & tic46096489 & 214.68301 & -20.27544 & toi818.01 & 12.719 & 3.119 \\ 
        3211188618762023424 & tic43647325 & 76.08189 & -6.22978 & toi423.01 & 14.788 & 3.162 \\ 
        2506859165273343488 & tic332715376 & 32.67117 & -0.67888 & toi4354.01 & 2.724 & 10.631 \\ 
        4617735638779366016 & tic318608749 & 9.12614 & -83.56707 & toi1100.01 & 4.138 & 20.073 \\ 
        45456357610673408 & epic210550063 & 61.13186 & 16.21665 & epic210550063.01 & 2.643 & 2.166 \\ 
        6357524189130820992 & tic317060587 & 337.51024 & -75.64656 & toi1052.01 & 2.434 & 9.14 \\ 
        4866555051425383424 & tic77253676 & 69.70423 & -36.68132 & toi697.01 & 2.527 & 8.608 \\ 
        4870809920906672384 & tic170729775 & 67.98437 & -34.45554 & toi2449.01 & 11.682 & 16.654 \\ 
        4877544322252002048 & tic1167538 & 70.99759 & -31.90664 & toi2447.01 & 10.005 & \nodata \\ 
\enddata
\end{deluxetable*}

\begin{figure*}
    \centering
    \includegraphics[width=\textwidth]{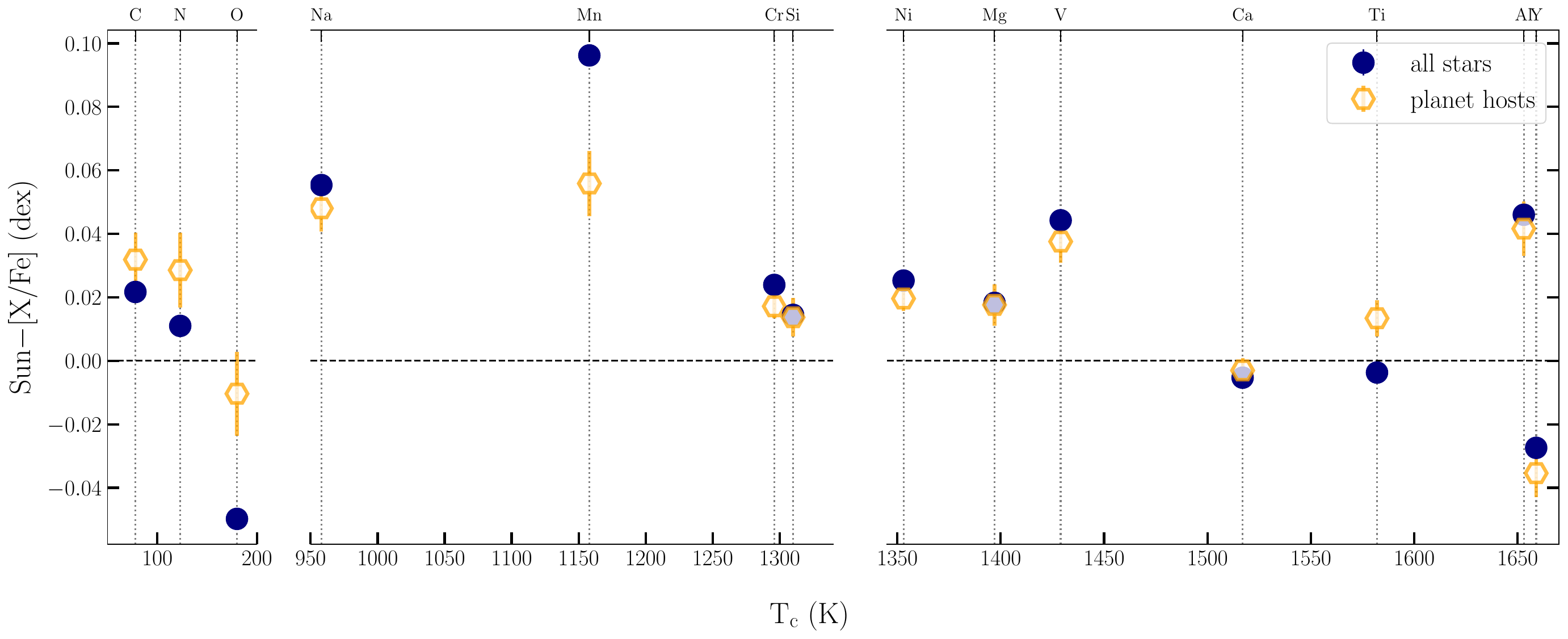}
    \includegraphics[width=\textwidth]{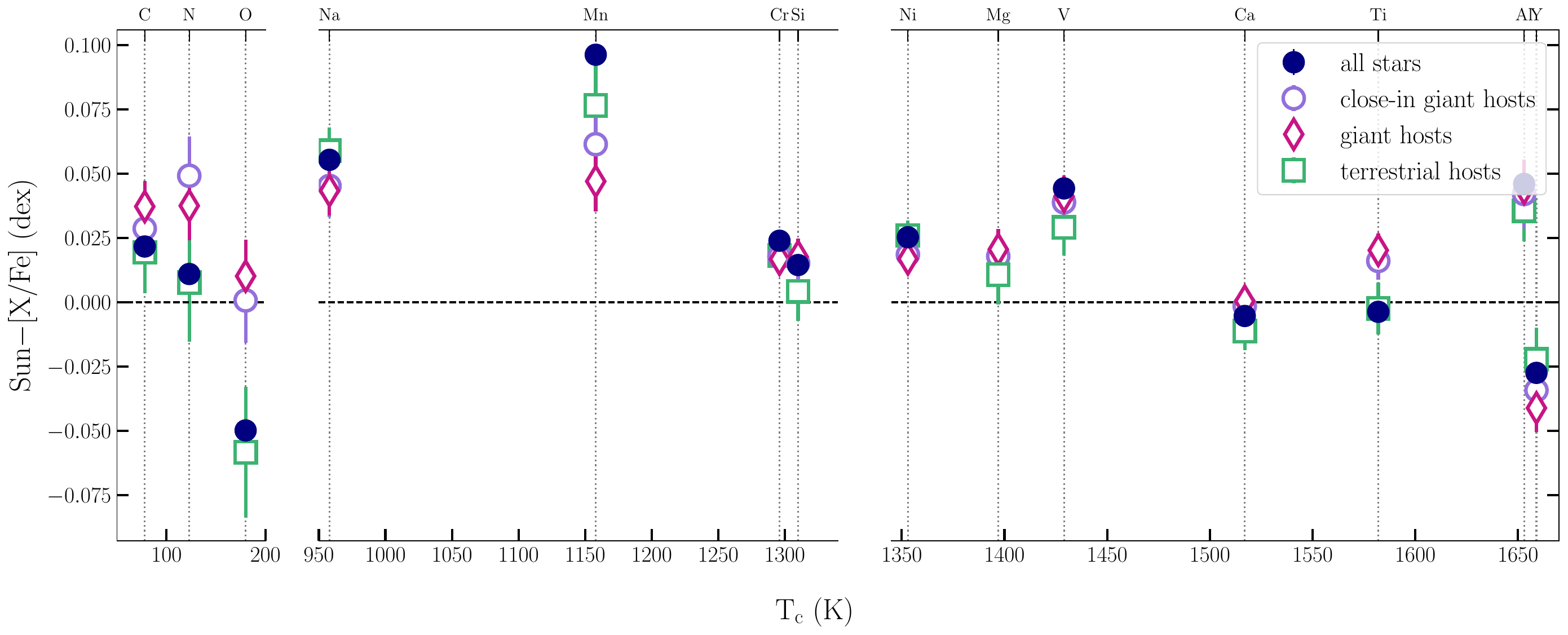}
    \caption{Mean abundance measurements of our sample with standard errors of the means (standard deviation divided by square root of number of stars) as a function of condensation temperature compared to the Sun. \textit{Top}: All test set solar analogs as navy filled circles and identified planet hosts as orange hexagons. The Sun appears relatively refractory depleted with increasing condensation temperature compared to these samples replicating \cite{bedell18}, and the solar analogs and planet hosts are chemically indistinguishable within 2-3$\sigma$. \textit{Bottom}: All test set solar analogs as navy filled circles, close-in giant planet hosts as purple unfilled circles, giant planet hosts as pink unfilled diamonds, and terrestrial/small planet hosts as green unfilled squares. The Sun still appears relatively depleted compared to these four samples, which are still chemically indistinguishable within 2-3$\sigma$.}
    \label{fig:bedell}
\end{figure*}

\begin{figure*}
    \centering
    \includegraphics[width=0.75\textwidth]{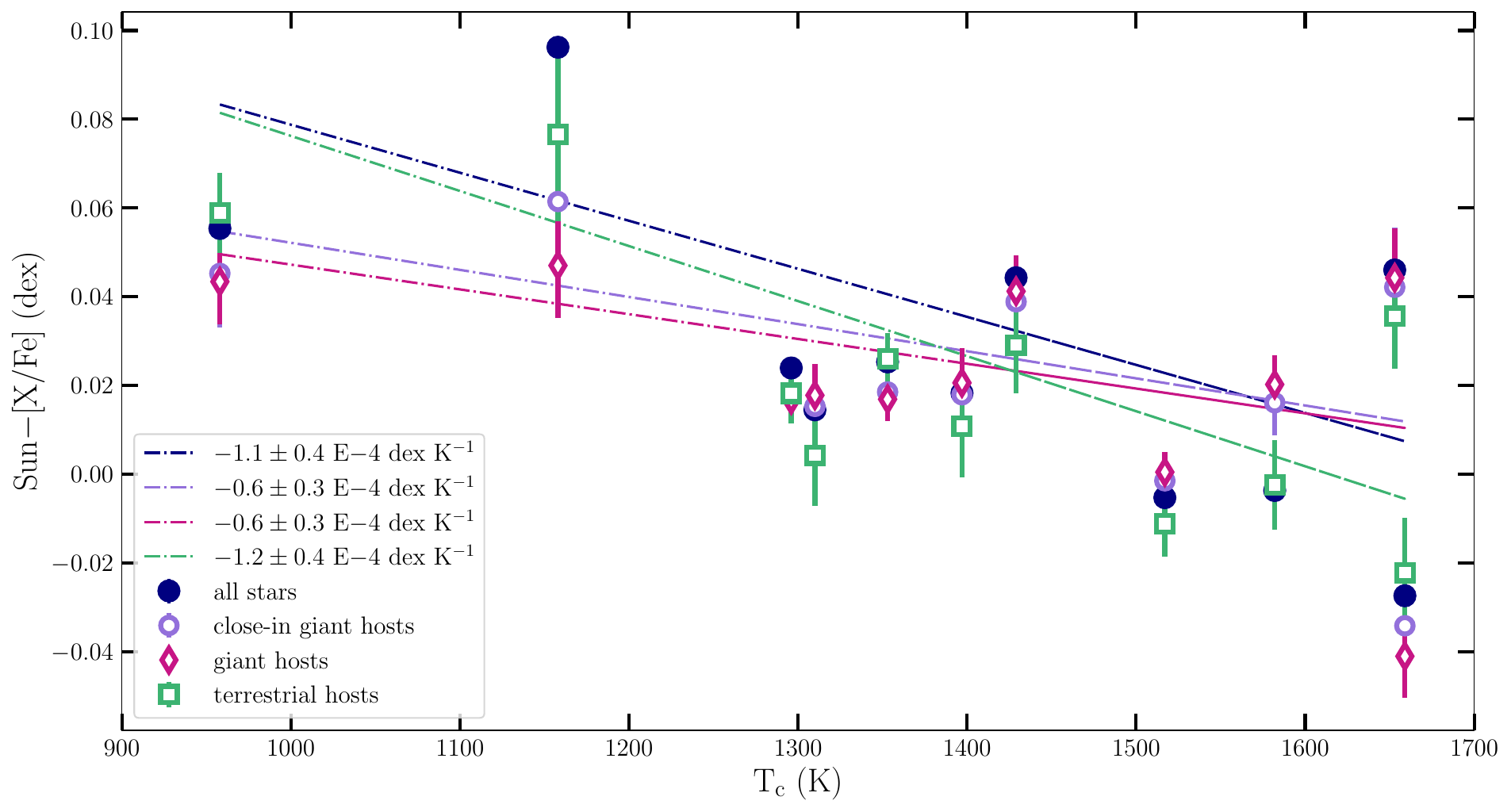}
    \caption{Linear fits of test set solar analogs (navy filled circles), close-in giant planet hosts (purple unfilled circles), giant planet hosts (pink unfilled diamonds), and terrestrial/small planet hosts (green squares) for element mean abundances and respective standard errors of the means from Na (958 K) to Y (1659 K). The slopes are indistinguishable within errors suggesting the increased relative depletion with condensation temperature of the Sun is unrelated to our current knowledge of planet host status.}
    \label{fig:bedell_ph}
\end{figure*}

\subsubsection{Refractory Depletion Trends for the Sun}
While measuring precise abundances for planet hosts has many applications, we examine the abundance space of solar analog planet hosts compared to the Sun. The Sun has long thought to be refractory depleted compared to other Sun-like stars. \cite{bedell18} confirmed this with an extensive high-resolution spectroscopic survey of solar twins, 80\% of which were refractory-enriched with increasing condensation temperature compared to the Sun. Abundance comparison efforts of solar analog planet hosts and the Sun suggested that the missing refractories are trapped by the terrestrial planets \citep{Melendez09,Ramirez09}, or the giant planets created dust traps thereby keeping refractories from the host star \citep{Booth20}. If either of these theories were true, we would expect to find that planet hosts with terrestrial planets or giant planets have the same abundances as the Sun.  

In Figure \ref{fig:bedell} we show the mean abundances and report the confidence on the mean with the error bars (as calculated by the standard deviation divided by the square root of the number of stars) compared to the Sun as a function of condensation temperature for our entire test set, all planet hosts, close-in giant planet hosts, giant planet hosts, and terrestrial planet hosts. The condensation temperatures used are the 50\% condensation temperatures calculated for solar system composition gas from \cite{Lodders03}. We examine the trend of refractory depletion with increasing condensation temperature starting with Na (958 K) to Y (1658 K). We also convert all of our abundances from [X/H] to [X/Fe] so our results are analogous to those of \cite{bedell18}. 

The entire test set does not significantly differ from any kind of planet host more than 3$\sigma$. In most cases, they are chemically indistinguishable within 1--2$\sigma$. We also see that all four populations are largely refractory enriched compared to the Sun replicating the finding of \cite{bedell18}. We show the results of fitting a line to each of these populations for just the refractory elements (Na, Mn, Cr, Si, Ni, Mg, V, Ca, Ti, Al, and Y) in Figure \ref{fig:bedell_ph}. Visually, it appears that the slope for all of the test set stars ($-1.1 \pm 0.4 \mathrm E-4$ dex K$^{-1}$) resembles the slope of the terrestrial planet hosts ($-1.2 \pm 0.4 \mathrm E-4$ dex K$^{-1}$). This is unsurprising as it has been shown that most sun-like stars likely host smaller planets \citep{Petigura13}. Similarly, the slopes for giant planet hosts and close-in giant planet hosts are almost exactly the same. This also make sense as there is a large overlap between the close-in giant planet and giant planet population in our sample. Astrophsyically, close-in giant planets could be from the same giant planet population but have just undergone type II migration \citep{Dawson18}. However, we find that all the slopes are indistinguishable within $1\sigma$ errors. These slopes are also on the same order of magnitude as those reported in \cite{bedell18}. We also measure slopes for each individual star compared to the Sun and find that the Sun is relatively refractory depleted compared to 87\% of our solar analog population (compared to Bedell's 80\%). This suggests that there is not a significant correlation between refractory enrichment or depletion and our current knowledge of planet host status; the Sun remains relatively refractory depleted compared to all types of planet hosts.   

\begin{figure*}
    \centering
    \includegraphics[width=\textwidth]{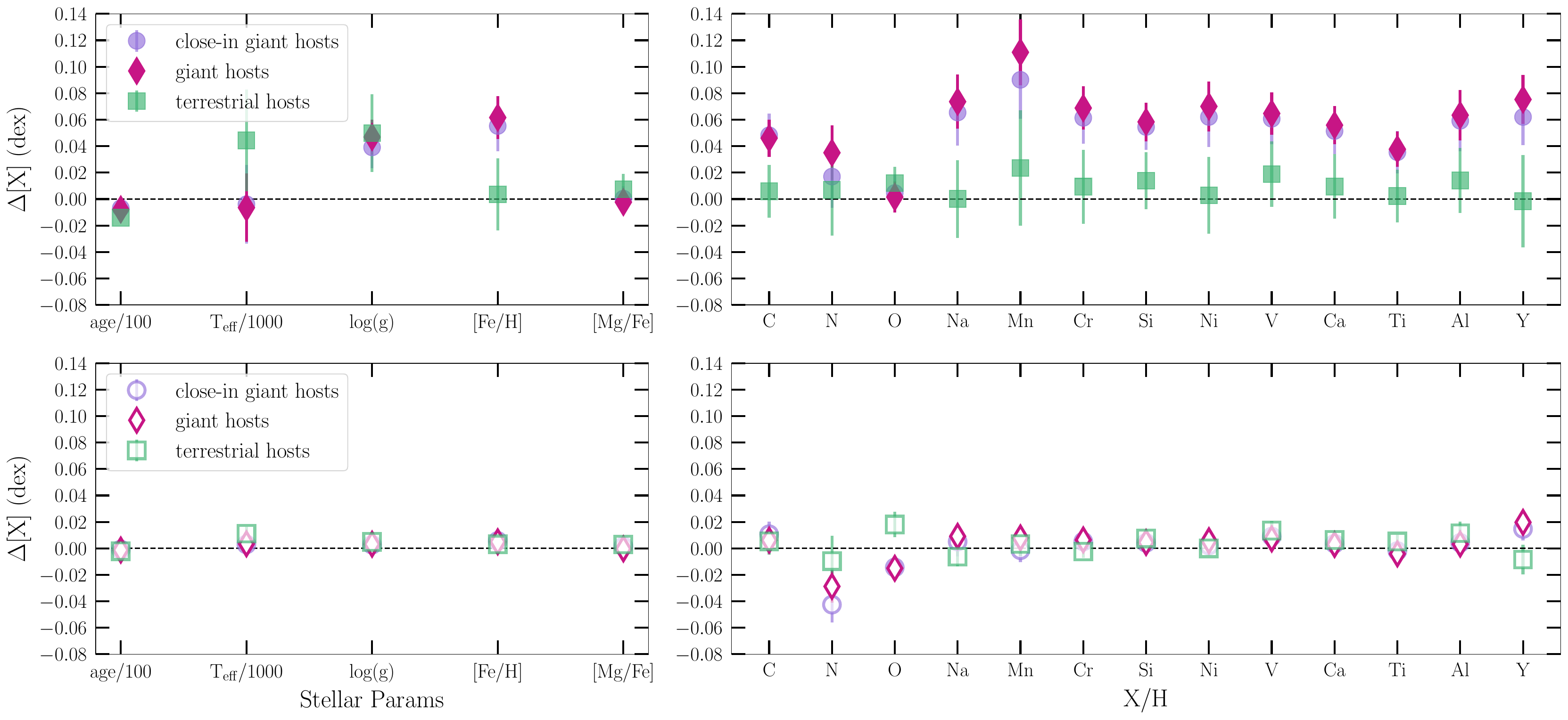}
    \caption{Mean label differences between close-in giant planet hosts (purple circles), giant planet hosts (pink diamonds), and terrestrial/small planet hosts (green squares) and non-planet hosts. \textit{Top:} There are differences in the five fundamental parameters (left) that could describe the differences seen in abundance space on the right, though no non-zero differences are more than 4$\sigma$. Errors on means are rms sum of standard deviation of planet hosts divided by square root of number of planet hosts plus standard deviation of test set stars divided by square root of number of test set stars. \textit{Bottom:} The five fundamental parameters have been ``doppelganger corrected'' (left). All identified planet hosts are compared solely against non-planet hosts with the same age, \teff, \logg, \feh, and \mg\ within errors. This decreases the spread in non-zero differences even further to $ < 2\sigma$. Error bars are standard error of the mean.}

    \label{fig:ph_dopps}
\end{figure*}

\subsubsection{Abundance Differences Between Planet Host Populations}\label{sec:doppel_test} 

Though the mean differences in abundances among the different types of planet hosts and the entire test set are already small, we investigate these differences further. We compare close-in giant planet hosts, giant planet hosts, and terrestrial planet hosts to the non-planet hosts (the other 17,362 in the test set). We show the mean differences in abundances between the planet hosts and non-planet hosts ($\Delta[\text{X}]$) in the top right panel of Figure \ref{fig:ph_dopps}. The terrestrial planet hosts are consistent in abundance space with the non-planet hosts. The close-in giant and giant planet hosts differ from the non-planet hosts, although no abundance has $ > 4\sigma$ difference. We also compare the mean differences for 5 fundamental stellar parameters: age, \teff, \logg, \feh, and \mg\ as the representative \alphafe\ element (from [Mg/H]--\feh). The results are shown in the top left panel. The non-planet hosts are consistent within $2\sigma$ for \teff, \logg, \feh, and \mg\ to terrestrial planet hosts; there is a $> 3\sigma$ offset of 0.01 dex in age. There are offsets $ > 3\sigma$ of 0.04--0.05 dex in \logg, and 0.06 dex in \feh\ for close-in giant and giant planet hosts. This raises the possibility that any differences in abundances between non-planet hosts and planet hosts could be driven by systematic population differences in age, \logg, and \feh\ rather than intrinsic abundance differences from planet architectures. We subsequently test and see this is the case, using a doppelganger comparison test.

We define doppelgangers to be stars that have the same fundamental parameters (at 1$\sigma$) as our test objects \citep{Behmard23,Sayeed23}, in labels of age, \teff, \logg, \feh, and \mg. This enables us to build a control sample to examine any differences in elemental abundances alone at otherwise fixed conditions of evolutionary state (age, \teff, \logg) and fiducial supernovae enrichment levels represented with Fe and \mg. For each close-in giant, giant, and terrestrial planet host, we find their associated doppelgangers in age, \teff, \logg, \feh, and \mg within $1\sigma$ from the non-planet host sample. We calculate the mean stellar parameter and abundance differences between each planet host minus their N randomly drawn doppelgangers\footnote{Because every planet host does not have the same number of doppelgangers, N is determined by the star with the fewest doppelgangers. This star will be compared to all of its doppelgangers, while other stars will be compared to a subset of their doppelgangers of size N.}. This gives us a representative distribution of the difference in each planet host and a doppelganger population. In the second row of Figure \ref{fig:ph_dopps}, we show the mean of these mean differences and standard error of this mean. 

The mean differences in stellar parameters (left bottom panel) between populations are of course zero since we are only comparing planet hosts to their doppelgangers. However, now we see that this in turn makes all previous non-zero abundance differences seen for planet hosts, particularly the close-in giant and giant planet hosts, consistent with zero within at most $2\sigma$. The mean difference for [N/H] is still nonzero post-doppelganger-correction, but recall this abundance was poorly labeled during the training step. The trend towards zero difference between populations agrees with those presented in \cite{Behmard23}, which also found a subset of planet hosts from the Kepler mission are chemically indistinguishable from their doppelgangers. 

We also inspect how other stellar parameters change for planet hosts when solely conditioning on \teff\ and \logg. We find that there are differences in age, \feh, and \mg\ that reflect overall abundance differences from star forming conditions as reflected by supernovae elements. We then iteratively condition on the three stellar parameters that did have systematic differences, age, \logg, and \feh, one at a time for all of the planet host populations. We find that conditioning on \feh\ has the largest impact in decreasing abundance differences. This makes sense given the well-known planet-metallicity correlation \citep{Fischer05}. In fact, we recover this exact finding in that the (close-in) giant planet hosts are on average more metal-rich than the non-planet and terrestrial planet hosts, which was causing the significant non-zero abundance differences in the top right panel of Figure \ref{fig:ph_dopps}. Conditioning on age and \logg\ also decreases the spread in abundance differences albeit at respectively smaller scales. Thus, conditioning on all fundamental stellar parameters simultaneously results in the greatest decrease in abundance differences. 


\section{Discussion}\label{sec:conclusions}
In this work, we use \cannon\ to train a linear model on 34 solar analog Gaia RVS stars with high precision literature stellar parameters and abundances, measured from high resolution data (R $> 30,000$), to infer \teff, \logg, and 15 individual abundances ([X/H]) for 17,412 other Gaia RVS (R$\approx 11,000$) stars, 50 of which are (candidate) planet hosts. This data-driven analysis yields average upper limit precisions of $61$ K in \teff, $0.09$ in \logg, and $0.04-0.1$ in [X/H] (see Table \ref{t:ts}). These are an improvement compared to the RVS measurement precisions, which are 70 K for \teff, 0.11 dex for \logg, and 0.15 dex for \feh\ compared to 0.036 dex from our results shown in Table \ref{t:intdisps}.  

In compiling the reference set, we used data from 5 different survey efforts \citep{Bensby14,Brewer16,Brewer18,bedell18,Hinkel14}. Our model works remarkably well despite typically seen offsets in literature labels due to different instruments and analysis methods used across these surveys. We see this in our initial cross-validation test described in Section \ref{sec:refset}. That is, when treating each star in the reference set itself as a test star and applying the model trained without that star, we can recover the literature element abundance values to precisions of $0.05 \pm 0.02$ dex (Figure \ref{fig:crossval_scatters}). \cannon\ also uses the inferred labels to generate model spectra that we can compare to the observed RVS spectra. We show an example spectrum from the cross-validation in Figure \ref{fig:genmodel_ex} and a summary of the residuals (data-model) of the \cannon-generated spectra and the RVS spectra for the test set in Figure \ref{fig:imshow}. Residuals are on average $< 1\%$.

\subsection{The Limits of Gaia RVS Inferred Abundances}
An important consideration when using data-driven methods for low to medium resolution surveys is what precision can be obtained. High-resolution spectrographs are advantagenous for spectroscopy due to the high precision measurements achieved. However, we are able to obtain surprisingly high precision at much lower resolution \citep{Ting2017}. Achieving high precision is relevant in the context of the abundance dimensionality. It appears that the dimensionality of the abundance space is low, and the many abundances we infer can be well predicted using age, \teff, \logg, and a small subset of elements. We find even just two element measurements ([Mg/H], \mg\ and \feh, \mg) are enough to predict the other elements, similar to \cite{Griffith21}. The intrinsic information in the elements beyond what information is captured in \teff, \logg, \feh, \mg, and age requires higher resolution and more precise observations to access, if it is present. We find we do not have the precision to measure any intrinsic scatter in the majority of the other 13 elements we derive when using our five label model to predict these abundances. Using just 2 labels, we can predict abundances to $> 87\%$, even after adjusting our error estimates from \cannon\ to account for the input reference label errors. Measuring potential and significant differences in abundances due to planet architectures therefore requires precisions higher than we achieve ($\lesssim 0.02$ dex), because we need to access information beyond what is already captured in the 2--5 labels. Additionally, the impact of time-variable stellar atmospheres and surface motions may add complexity to measuring accurate and precise abundances at this level \citep{Spina2020}. This level of measurement precision for element abundances is more readily achievable with Gaia RVS spectra of S/N $\gtrsim 200$ (see Figure 4 of \citealt{Ting2017}), which only includes 2\% of our current sample though this can be reached in future data releases after spectra are co-added with multiple visits. This precision is certainly more accessible with high resolution spectroscopy (R $>$ 100,000). 

While we have 44 stars in our test set with the same abundances and stellar parameters as the Sun within their errors, we have not necessarily found 44 solar birth siblings. The prospect of finding stellar birth siblings, which presumably have identical abundances, also requires this extremely high precision, and it may also be out of reach even at abundance precisions of $\sim$0.02 dex \citep[e.g.][]{Ness_2018}. Using the potential differentiating power of abundances also requires measuring elements from multiple nucleosynthetic families \citep[e.g.][]{Freeman2002, Manea23}.

We infer [X/H] for the majority of the elements. \cite{bedell18} and \cite{Adibekyan16} discuss the merits of using [X/Fe] versus [X/H], which we consider here. [X/Fe] represents the enhancement or depletion of a given element from the norm of stars for a given bulk metallicity, \feh, and can mitigate trends related to Galactic chemical evolution compared to [X/H]. [X/H] represents the elemental abundance with respect to the production of elements in the big bang. We choose to use [X/H] as the most direct transfer of labels from the training set. We test how well \cannon\ performs using [X/Fe] versus [X/H] and find that \cannon\ has a poorer performance and a higher bias for many labels.

It is also important to recognize the differences between abundance \textit{inferences} and physical abundance measurements. In particular, C, O, Na, V, Al, and Y absorption lines are not present in the RVS wavelength region, which would typically prevent measurements of these elements in traditional abundance analysis. However, \cannon\ can use other lines to infer abundances for these elements (see the coefficients in Appendix \ref{sec:modelcoeffs}). Additionally, some elements such as O produce variations across the entire spectrum due to electron donorship (e.g. \citealt{Ting18}), so in a number of cases it is appropriate to use the full spectrum to infer an element, not simply the specific absorption lines. 

\subsection{The Sun Remains Relatively Refractory Depleted}
 Our finding that the element abundances between the populations of identified planet hosts and other solar analog stars are indistinguishable can be used to help interpret the Sun's relative refractory depletion. The Sun has been shown to be relatively depleted in refractory elements with increased condensation temperature compared to 80\% of other Sun-like stars \citep{bedell18}. In our work, we also find this trend holds true for the Sun compared to 87\% of our solar analog test set regardless of planet host status (Figures \ref{fig:bedell} and \ref{fig:bedell_ph}). The same trend holds true for the 382 most solar-like stars in our test set with the same age, \teff, \logg, \feh, and \alphafe\ as the Sun within their errors. Thus, the source of this apparent refractory depletion trend in the Sun remains an open question. 

Early work such as \cite{Melendez09,Ramirez09} suggests that refractory depletion trends can change depending on planet host type. More specifically stars like the Sun, with terrestrial planets, show this trend because terrestrial planets lock up the refractories. Such planet hosts will appear refractory depleted compared to those hosting close-in giant planets. This suggestion has been challenged by the modeling work of \cite{Kunitomo18} who shows $ > 16$ M$\oplus$ of refractory rich material far exceeding that of the terrestrial planets is needed to make up the deficit seen in the Sun. Our work also challenges this theory empirically given that we see, regardless of type of planet host, the Sun still remains relatively refractory depleted with increasing condensation temperature in comparison. What was likely being seen in \cite{Melendez09,Ramirez09} were small but systematic differences in fundamental stellar parameters among the identified planet hosting populations that propagated to abundance differences rather than intrinsic abundance differences.

Our results, given the precision limits, are inconsistent with the theory suggested by \cite{Booth20}. They posit that giant planets, like Jupiter, could create dust traps that sequester refractory-rich dust and prevent it from falling into the host star. This has been suggested as an explanation for the abundance differences and measured refractory depletion trend of $\approx 5\mathrm E -5$ dex K$^{-1}$ in the wide binary system HD196067-HD196068 \citep{Flores24}, which is similar to the slope we measure for giant planet hosts ($6\mathrm E -5$ dex K$^{-1}$, Figure \ref{fig:bedell_ph}). However, we see in the top panel of Figure \ref{fig:ph_dopps} that giant planet hosts (close-in or not) are chemically indistinguishable within errors from stars that are not known giant planet hosts. \cite{Huhn23} recently calculated that abundance differences due to dust traps from giant planets are $\approx 0.01$ dex. So while giant planets could sequester elements, it is not happening at a precision we are necessarily able to detect. The solar analogs in this work \textit{do} appear enriched compared to the Sun at scales larger than those shown in \citealt{Huhn23} across all planet host types. Thus, the source of the Sun's refractory depletion seems to not only be from the planets it currently hosts (if it is at all). 

There are obvious biases in our planet sample due to detection incompleteness. This is particularly true for smaller radius and long-period in spite of the close-in giant planets in our sample having plausibly experienced type II migration from the outer part of their disks. We present these results with this caveat and do not outright refute any theories of planets suppressing refractories. Rather, we show there is no significant link between planet host status and stellar refractory composition emerging based on our current observations.



From our work, we calculate that the Sun is in the 13th percentile of the solar analog abundance pattern distribution. The planet hosts we study are likely sampled from this abundance distribution. We confirm this is true by randomly drawing 43 stars from our test set that are \textit{not} the 43 identified planet hosts 100 times. We measure the slope of their mean refractory abundances as a function condensation temperature and find an average slope of $-1.1 \pm 0.2 \mathrm E -4$ dex K$^{-1}$. This is in agreement with results from our linear fits shown in Figure \ref{fig:bedell_ph}. This indicates that the Sun is indeed a bit lower in the general sun-like star abundance pattern distribution but is still reasonably part of a continuous distribution. Therefore, the refractory element trends may not actually require any special explanations.


\subsection{Prospects for Exoplanet Composition \& Formation Studies}
This catalog is a demonstration of the utility of data-driven learning in the era of big data surveys. There are certainly limits to the precision and dimensionality of Gaia RVS data. However, we have been able to infer abundances for 17,412 stars and 50 planet (candidate) hosts using a reference set made up of stars from different high resolution spectroscopic surveys. 

Host star abundances can play a role in constraining exoplanet compositions and formation pathways. While the compositions of planets can span a wider range than that of their host stars \citep{Plotnykov20}, rocky planets have been shown to mirror [Fe/Mg] and [Si/Mg] ratios of their host stars. In our own solar system, Earth and Mars have [Fe/Mg] and [Si/Mg] ratios that match the Sun's to within 10\% \citep{Unterborn19}. \cite{Rodriguez23a} finds this is true of K2-106b, and \cite{Adibekyan21} generally finds a strong correlation (although not 1:1) between planet iron mass fraction and host star composition. This underscores the importance of refractory elements in rocky planet structure. We infer abundances for Si, Fe, Mg, and 9 other refractory elements. %
Abundance measurements of volatile elements like C and O, which we also infer, are also useful as they can trace the formation and migration of planets using the various molecular ice lines present in the protoplanetary disk \citep{Oberg11}. In order to differentiate between planet populations among stars, \cite{Hinkel18} calculate that precisions of 0.01 for [Si/H] and 0.02 for \feh\ are needed. The precision only increases for other elements. 
Our average upper limit precisions for Si and Fe are 0.04 dex. These precisions improve to 0.02 and 0.03 dex if we account for input reference label errors. While this entire catalog may not be used to infer or differentiate between planet compositions just as yet, we have used these data to understand the extent of the Sun's apparent refractory depletion compared to $ > 17,000$ solar analogs. With near all sky surveys like Gaia, data-driven methods like \cannon, and increased future exoplanet discoveries from the upcoming Plato Mission \citep{Rauer14} and the Nancy Grace Roman Space Telescope \citep{Spergel15}, we can assess such abundance and stellar parameter trends for stars hosting planets on a much larger and more robust scale.



\acknowledgments

We thank the anonymous referee for the extremely quick turnaround and constructive comments that helped us improve the manuscript. We thank the THYME Collaboration, Em Boudreaux, and Aylin Garcia Soto for helpful discussions. RR thanks the LSSTC Data Science Fellowship Program, which is funded by LSSTC, NSF Cybertraining Grant \#1829740, the Brinson Foundation, and the Moore Foundation; her participation in the program has benefited this work.

This work has made use of data from the European Space Agency (ESA) mission Gaia (https://www.cosmos.esa.int/gaia), processed by the Gaia Data Processing and Analysis Consortium (DPAC, https://www.cosmos.esa.int/web/gaia/dpac/consortium). Funding for the DPAC has been provided by national institutions, in particular the institutions participating in the Gaia Multilateral Agreement. This research has made use of the NASA Exoplanet Archive, which is operated by the California Institute of Technology, under contract with the National Aeronautics and Space Administration under the Exoplanet Exploration Program.


%

\vspace{5mm}
\facilities{Gaia \citep{gaiamission}, NASA Exoplanet Archive \citep{ps}}


\software{astropy \citep{2013A&A...558A..33A,astropyii}, sklearn \citep{sklearn}, scipy \citep{scipy}, Cannon \citep{cannon}, Korg \citep{KorgI,KorgII}, numpy \citep{numpy}, pandas \citep{reback2020pandas,mckinney-proc-scipy-2010}, topcat \citep{topcat}, isoclassify \citep{Huber17,Berger20}, matplotlib \citep{Hunter:2007}, bayestar19 \citep{Green3D2019}
          }

%

\bibliographystyle{aasjournal}


\appendix 
\renewcommand\thetable{\thesection.\arabic{table}}
\renewcommand\thetable{\thesection.\arabic{figure*}}
\renewcommand{\thesubsection}{\Alph{subsection}}
\counterwithin{figure}{section}
\counterwithin{table}{section}
\setcounter{figure}{0}

\section{\textbf{Model Coefficients}}\label{sec:modelcoeffs}
\begin{figure*}[h!]
    \centering
    \includegraphics[height=0.82\textheight]{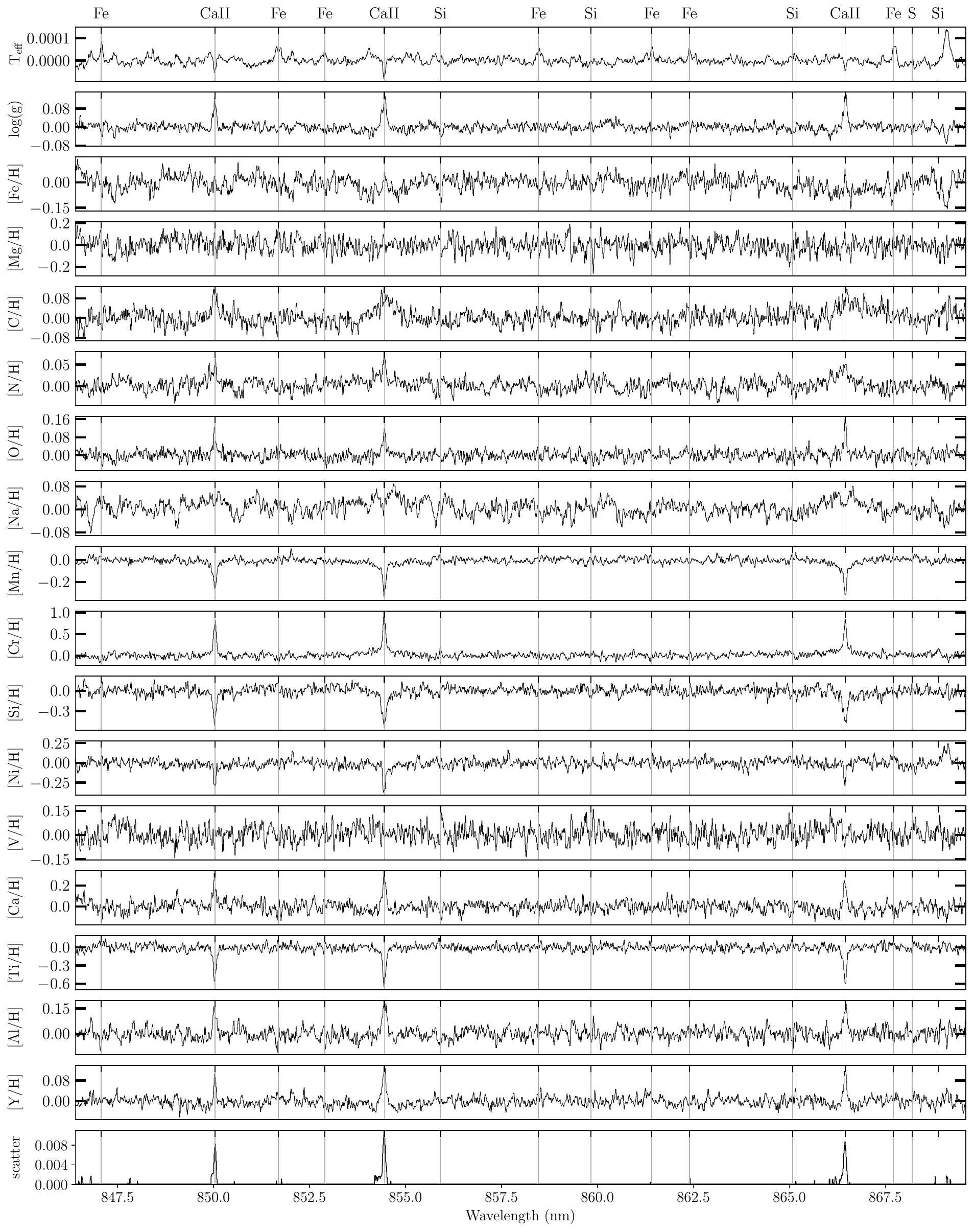}
    \caption{Coefficients of \cannon-model for each label. Most labels are best ``learned'' from the variance in the cores and wings of the Ca II lines, which is consistent with \cite{Rampalli21}.}
    \label{fig:coeffs}
\end{figure*}




\end{document}